\documentclass[useAMS,usenatbib]{mn2e}
\usepackage{graphicx}
\usepackage{mathptmx}
\usepackage{dcolumn}
\newcolumntype{d}{D{.}{.}{-1}}

\def\SNR(#1.#2)#3(#4.#5){{G#1${\cdot}$#2$#3$#4${\cdot}$#5}}
\graphicspath{{.}}
\newcounter{todo}
\renewcommand\thetodo{\Alph{todo}}
\def\todo#1{\addtocounter{todo}{1}[[\thetodo: #1]]\strut\vadjust{%
\kern-\dp\strutbox{\vtop to \dp\strutbox{\baselineskip\dp\strutbox\vss\rlap{%
\hskip\hsize\ \rm{$\leftarrow$\thetodo}}\null}}}}
\def\note#1{\strut\vadjust{\kern-\dp\strutbox{\vtop to \dp\strutbox{%
\baselineskip\dp\strutbox\vss\rlap{\hskip\hsize\ {\tiny\rm #1}}\null}}}}
\title[AMI observations of northern supernova remnants at 14--18\,GHz]{AMI
observations of northern supernova remnants at 14--18\,GHz\thanks{We request that any reference to this paper cites ``AMI Consortium: Hurley-Walker et al. 2009"}}

\label{firstpage}

 \author[Hurley-Walker et~al.]{AMI Consortium:
 Natasha~Hurley-Walker\thanks{Issuing author: email -- nh313@mrao.cam.ac.uk}, A.~M.~M.~Scaife,
 D.~A.~Green,\newauthor
 Matthew~L.~Davies, Keith~Grainge,
 Michael~P.~Hobson, Michael~E.~Jones$\ddagger$, \newauthor
 Tak~Kaneko, Anthony~Lasenby, Guy~Pooley, Richard~D.~E.~Saunders,
 Paul~F.~Scott,\newauthor
 David~Titterington, Elizabeth~Waldram and Jonathan~T.~L.~Zwart\\
Astrophysics Group, Cavendish Laboratory, Cambridge University,
19 J.~J.~Thomson Avenue, Cambridge CB3 0HE\\
$\ddagger$ Astrophysics, Oxford University, Keble Road, Oxford OX1 3RH
}

\date{Accepted ---; received ---; in original form \today}

\pagerange{\pageref{firstpage}--\pageref{lastpage}}

\pubyear{2008}

\begin{document}

\maketitle

\begin{abstract}
We present observations between 14.2 and 17.9\,GHz of 12 reported supernova remnants
(SNRs) made with the Arcminute Microkelvin Imager Small Array (AMI SA). In conjunction with
data from the literature at lower radio frequencies, we determine spectra of
these objects. For well-studied SNRs (Cas A, Tycho's SNR, 3C58 and the Crab Nebula),
the results are in good agreement with spectra based on previous results.
For the less well-studied remnants the AMI SA observations provide higher-frequency radio
observations than previously available, and better constrain their radio spectra.
The AMI SA results confirm a spectral
turnover at $\simeq 11$\,GHz for the filled-centre remnant \SNR(74.9)+(1.2). We also see a possible steepening of the spectrum of the filled-centre remnant \SNR(54.1)+(0.3) within the AMI SA frequency band
compared with lower frequencies.
We confirm that \SNR(84.9)+(0.5), which had previously been identified as a SNR,
is rather an {\sc Hii} region and has a flat radio spectrum.

\end{abstract}


\section{Introduction}

We present observations of 12 reported supernova remnants (SNRs) selected from the 2006 version of
Green's SNR catalogue\footnote{http://www.mrao.cam.ac.uk/surveys/snrs/;
see also \citet{2004BASI...32..335G}.} made with the Arcminute Microkelvin
Imager Small Array (AMI SA, see \citealt{2001MNRAS.328..783K} and \citealt{0807.2469}) at 14.2 to 17.9\,GHz. The
integrated flux densities from the AMI SA observations are used in conjunction with
measurements from the literature to investigate the synchrotron
 spectrum of these remnants.
There are few observations of Galactic SNRs at frequencies above
10\,GHz, and several issues can be addressed by observing at higher frequencies. Generally,
observations over a wide range of frequencies are essential to investigate energy
injection and loss processes in the intergalactic medium.
More specifically, `filled-centre' remnants have relatively flat,
non-thermal spectra, with spectral indices $\alpha$ typically between 0.0 and 0.3 (here
we define $\alpha$ such that flux density $S$ varies with frequency $\nu$ as
$S\propto\nu^{-\alpha}$). Such flat spectra steepen at high frequencies;
the positions of features such as spectral breaks, when known, vary considerably between
different remnants \citep[e.g.][]{1987A+AS...71..189M,1989ApJ...338..171S, 1992MNRAS.258..833G,2005ApJ...626..343B}. In the case of several filled-centre
remnants the frequency of the turnover is not yet known. Also, there have been
suggestions (\citealt{1992ApJ...399L..75R, 2005A+A...436..187T,
2007ApJ...655L..41U}, see also \citealt{2006A+A...451..991T}) that `shell' type
remnants, which typically have $\alpha$ in the range 0.3 to 0.7, show spectral
flattening at higher frequencies. However, the detection of small changes in
spectral index over a limited range of frequencies is not easy
(see e.g. \citealt{2007BASI...35...77G}), and observations over a wider range of frequencies
is an advantage in such studies. Finally, possible excess `anomalous' emission
-- which may be from spinning dust -- has been reported in the case of one
Galactic SNR (3C396, \citealt{2007MNRAS.377L..69S}). These AMI SA observations
provide additional constraints on any similar emission associated with the
observed remnants.

\section{The Telescope}\label{sec:telescope}

The AMI SA is situated at the Mullard Radio
Astronomy Observatory, Cambridge. It consists
of ten 3.7-m-diameter equatorially-mounted dishes with a baseline range of
$\simeq 5$--20\,m. The telescope observes in the band 12--18\,GHz with
cryogenically-cooled indium-phosphide HEMT front-end amplifiers. The system temperature is typically
about 25~K. The astronomical signal is mixed with a 24-GHz local-oscillator signal
to produce an intermediate-frequency signal of 6--12\,GHz. The correlator is an analogue Fourier
transform spectrometer with 16 correlations formed for each baseline at path
delays spaced by approximately 26~mm. From these correlations, the complex signals in each of eight
channels of 750~MHz bandwidth are synthesized. Note that for correlators of this
type, the synthesized channels are correlated at the $\simeq 10$ per cent level. In practice, the
lowest two frequency channels are presently unused due to a low response in
this frequency range, and interference from geostationary satellites.
The FWHM of the
primary beam of the AMI SA is $\approx 20$\arcmin at 16\,GHz and the FWHM of its synthesised beam
is at minimum $\simeq 2$\arcmin. The average synthesized beam
for a map made with combined frequency channels is shown in the figures presented in this paper, and is an effective measure of the resolution.

\section{Observations}

Observations of 12 reported SNRs (see Table~\ref{tab:srclist}) were made with the AMI
SA during the period June--July 2007. These targets were selected from
Green's SNR catalogue on the basis of angular diameter ($\le 12$\arcmin, to be mapped in a single pointing) and declination
 ($\delta > 18^\circ$, to be accessible, given the hardware constraints at the time).
The sample of SNRs includes several bright, well-studied
objects in the Galactic anti-centre: \SNR(111.7)-(2.1) ($=$Cas~A),
\SNR(120.1)+(1.4) ($=$Tycho's SNR), \SNR(130.7)+(3.1) ($=$3C58) and
\SNR(184.6)-(5.8) ($=$Crab Nebula). Comparison of AMI SA maps, flux densities and spectra of these objects
with data from the literature provides a useful verification of the performance
of the telescope.
The sample also includes less well studied objects, and indeed one has recently been
revealed by \citet{2007ApJ...667..248F} as an {\sc Hii} region rather than a SNR (see further discussion of
\SNR(84.9)+(0.5) below).
The positions and effective thermal noise associated with each AMI SA observation
are shown in Table~\ref{tab:srclist}; note that the noise levels for the brighter objects are limited by the dynamic range available.
The observations were typically eight hours long
and used interleaved observations of bright, nearby point sources at hourly intervals
for phase calibration. The sensitivity of AMI SA over all baselines, all channels, is $\approx 30~{\rm
mJy~s^{-1/2}}$ giving $\approx 0.2$\,mJy after eight hours. Given that
the objects observed here are bright, eight hours is more than sufficient to
obtain accurate flux densities; however the long observations are important to produce
good maps, whose structure is dependent on the $uv$ coverage.

\begin{table*}
\centering
\caption{Our SNR sample. Names are of the form G{[}galactic longitude]$\pm${[}galactic latitude].
Column 4 contains the effective thermal
noise for the AMI SA observation of each object. Column 6 denotes the type of SNR:
S = shell-type, F = filled-centre, ? = unclear. Positions, angular sizes and
types are from Green's catalogue, with angular size for \SNR(59.5)+(0.1) corrected
(see text).\label{tab:srclist}}
\begin{tabular}{lcccccc}\hline
 Name  & $\alpha$  & $\delta$  & $\sigma_{\rm th}$  & $\theta$ & Type & Common\\
       & (J2000) & (J2000) & (mJy~beam$^{-1}$) & arcmin &  & name \\ \hline
  \SNR(54.1)+(0.3) & 19 30 31 & +18 52 & 0.52 &    1.5      & S  &  \\
  \SNR(57.2)+(0.8) & 19 34 59 & +21 57 & 0.17 &    12       & S? &  \\
  \SNR(59.5)+(0.1) & 19 42 33 & +23 35 & 0.48 &     5       & S  &  \\
  \SNR(63.7)+(1.1) & 19 47 52 & +27 45 & 0.19 &     8       & F  &  \\
  \SNR(67.7)+(1.8) & 19 54 32 & +31 29 & 0.22 &     9       & S  &  \\
  \SNR(74.9)+(1.2) & 20 16 02 & +37 12 & 0.84 & $8\times6$  & F  &  \\
  \SNR(76.9)+(1.0) & 20 22 20 & +38 43 & 0.17 & $12\times9$ & ?  &  \\
  \SNR(84.9)+(0.5)$^{a}$ & 20 50 30 & +44 53 & 0.32 &     6       & S  &  \\
 \SNR(111.7)-(2.1) & 23 23 26 & +58 48 & 92.0  &     5       & S  & Cas A \\
 \SNR(120.1)+(1.4) & 00 25 18 & +64 09 & 2.9  &     8       & S  & Tycho's SNR \\
 \SNR(130.7)+(3.1) & 02 05 41 & +64 49 & 14  & $9\times5$  & F  & 3C58  \\
 \SNR(184.6)-(5.8) & 05 34 31 & +22 01 & 165   & $7\times5$  & F  & Crab Nebula \\ \hline
\end{tabular}
\begin{minipage}{16cm}

 $^{a}${\sc Hii} Region; see text.

\end{minipage}
\end{table*}
\begin{table}
\centering
\caption{Assumed I~+~Q flux densities of 3C286 and 3C48, and errors on flux measurements in each frequency channel, over the commonly used AMI SA bandwidth.  \label{tab:Fluxes-of-3C286}}
\begin{tabular}{ccccc}\hline
 Channel & $\bar{\nu}$/GHz & $S^{{\rm {3C286}}}$/Jy & $S^{{\rm {3C48}}}$/Jy & $\sigma_{\rm {S}}$ \\ \hline
 3 & 14.24 & 3.61 & 1.73 & 3.0\% \\
 4 & 14.96 & 3.49 & 1.65 & 2.3\% \\
 5 & 15.68 & 3.37 & 1.57 & 1.9\% \\
 6 & 16.41 & 3.26 & 1.49 & 2.1\% \\
 7 & 17.13 & 3.16 & 1.43 & 1.7\% \\
 8 & 17.86 & 3.06 & 1.37 & 3.4\% \\ \hline
\end{tabular}
\end{table}
\begin{table*}
\centering
\caption{Phase calibrators used during SNR observations. 16-GHz flux densities were measured with AMI SA. \label{tab:Phase-calibrators-used}}
\begin{tabular}{lccccccc}\hline
 Source & $\alpha$~(J2000) & $\delta$~(J2000) & $S_{16}^{{\rm {AMI SA}}}$/Jy & \textsc{SNR} calibrated\\ \hline
 J1925+2106 & 19 25 59.6 & +21 06 26 & 2.15 &
 \SNR(54.1)+(0.3), \SNR(57.2)+(0.8),\\
 &&&& \SNR(59.5)+(0.1), \SNR(63.7)+(1.1)\\
 J2007+404 & 20 07 44.9 & +40 29 49 & 2.09 & \SNR(67.7)+(1.8),
 \SNR(74.9)+(1.2),\\
 &&&& \SNR(76.9)+(1.0)\\
 J2052+365 & 20 52 52.1 & +36 35 35 & 1.17 & \SNR(84.9)+(0.5)\\
 J2355+498 & 23 55 09.5 & +49 50 08 & 1.01 & \SNR(111.7)-(2.1)\\
 J0019+734 & 00 19 45.8 & +73 27 30 & 2.05 & \SNR(120.1)+(1.4)\\
 J0217+738 & 02 17 30.8 & +73 49 33 & 3.38 & \SNR(130.7)+(3.1)\\
 J0530+1331 & 05 30 56.4 & +13 31 55 & 4.53 & \SNR(184.6)-(5.8)\\ \hline
\end{tabular}

\end{table*}

\section{Calibration and Data reduction}\label{sec:data}

Data reduction was performed using the local software tool \textsc{reduce},
developed from the VSA data-reduction software of the same name. This applies
appropriate path compensator and path delay corrections, flags interference,
shadowing and hardware errors, applies phase and amplitude calibrations and
Fourier transforms the correlator data readout to synthesize the frequency channels,
before output to disk in $uv$ FITS format suitable for imaging in \textsc{aips}.

Flux calibration was performed using short observations of 3C48 and 3C286 near
the beginning and end of each run, with assumed I~+~Q flux densities for these sources in
the AMI SA channels consistent with \citet{1977A+A....61...99B} (see Table~\ref{tab:Fluxes-of-3C286}).
As \citet{1977A+A....61...99B} measure I and AMI SA measures I~+~Q, these flux densities
include corrections for the polarization of the sources derived
by interpolating from VLA 5-, 8- and 22-GHz observations.

Secondary calibrators were selected from the Jodrell Bank VLA Survey
(JVAS; \citealt{1992MNRAS.254..655P, 1998MNRAS.293..257B,
1998MNRAS.300..790W}) on the basis of their declination and flux density (see Table~\ref{tab:Phase-calibrators-used}). Over one
hour, the phase is generally stable to $5^{\circ}$ for channels 4--7, and
$10^{\circ}$ for channels 3 and 8.

The system temperature is
continously monitored using a modulated noise signal injected at each
antenna; this `rain gauge' is used to continuously correct the amplitude
scale in a frequency-independent way. The overall consistency of the
flux density scale is estimated to be better than 5 per cent.

The reduced visibility data were imaged using \textsc{aips}. Maps were made
from both combined channel datasets -- for channels 3 to 8 inclusive -- shown in this paper, and from individual
channels. The broad spectral coverage of AMI SA allows a representation of the
spectrum between 14 and 18\,GHz to be made.

Since the AMI SA antennas are sensitive to a single linear polarization
(I+Q) and are equatorially mounted,
this polarization is fixed on the sky during the observation;
therefore, the intensity of polarised sources may be
under- or overestimated. However, for the integrated flux densities,
this is not expected to be more than 10 per cent, which is comparable to the
quoted (statistical) uncertainties. For example, \citet{1966ApJ...144..437B}
obtain integrated percentage polarizations of $\simeq 9$ per cent and
$\simeq 1$ per cent at 15\,GHz for the Crab Nebula and Cas A respectively.
While some SNR have large polarization on small scales,
the AMI SA observations integrate over the whole source so this effect is reduced.
Errors on the AMI SA data points were estimated by combining the contributions
of several sources of error:
\begin{itemize}
\item Thermal noise on each channel and the dynamic range limit give rms map noise $\sigma_{\rm{rms}}$;
\item error on flux calibration (including rain gauge correction) of $\simeq 3$ per cent (see Table~\ref{tab:Fluxes-of-3C286});
\item we increase the flux calibration error to $10$ per cent in the case of those sources which undergo flux loss correction;
\item estimate of flux using tilted-plane method: changes in fitting area result in $\leq 1$ per cent changes in amplitude.
\end{itemize}
Thus the overall error was estimated as $\sigma =
\sqrt{\left(\sigma _{\rm{rms}}^{2}+\left[0.03~\rm{or}~0.10\right]S_{\rm{i}}\right)^{2}}$.

Spectra were compiled from the flux densities presented here in conjunction
with those available in the literature. Power-law spectra were fitted to these
data using a Gaussian likelihood function and Markov Chain Monte Carlo (MCMC) \citep{1953JoCP...21.1087}
sampling technique. This method is much faster than the traditional
parameter estimation methods and provides an error estimate on the spectral index directly
from the posterior parameter distribution. The resulting spectral indices are given in Tables~\ref{tab:oldsnrflux} and \ref{tab:newsnrflux}.

The correlation between AMI SA channels mentioned in Section~\ref{sec:telescope} is internal to AMI SA dataset, so its effect
on the spectral fit to AMI SA data and that from the literature is minimal.

\section{Results and Discussion}

Deriving accurate flux densities for extended objects in the Galactic plane, such as SNRs,
is not straightforward. In the majority of remnants presented here we
have adopted the fitting method of \citet{2007BASI...35...77G}. In this method
a flux density is estimated by drawing a polygon around the SNR and fitting a
tilted plane to the pixels around the edges of the polygon. The tilted plane
is then removed from the image before integrating the emission within the
polygon. Where there is a bright source closely adjacent to the SNR, such as in
the case of \SNR(74.9)+(1.2), the dividing line between the SNR and adjacent source is
omitted in fitting the plane. It was found that the flux density varied by less
than 1 per cent with different fitting regions; a contribution for this error is included
in the error on each flux density derived using this method.

As an interferometric telescope AMI SA does not measure total power and
therefore will lose flux on different angular scales depending on the
individual $uv$ coverage towards each source. Where total power maps
are publically available, it is possible to quantify the amount of flux
loss in the AMI SA data, or indeed any interferometric data, subject to
the assumption that the morphology of the region is unchanged between the
frequency of the total power map and that of the interferometer. In this
paper the flux loss inherent in each AMI SA channel towards the sources
\SNR(67.7)+(1.8), \SNR(74.9)+(1.2), \SNR(76.9)+(1.0), \SNR(84.9)+(0.5), \SNR(120.1)+(1.4), and \SNR(130.7)+(3.1) has
been calculated by sampling the Fourier transform of the total power maps from the Canadian Galactic Plane Survey (CGPS;
\citealt{2006A+A...457.1081K}) archives at 1.4\,GHz to match the $uv$ coverage of the corresponding AMI
observation. The remainder of our SNR sample
is not yet covered by the publically available CGPS data and the same method
cannot be applied. In the case of \SNR(63.7)+(1.1), where we believe flux loss
to be important, we have calculated a percentage loss for each channel using a
Gaussian model for the source and sampling it appropriately.
In the cases of
those sources that have not been corrected for flux loss, the importance of
this is discussed in the notes on individual sources.

\subsection{Well-studied SNR}
There are several SNR in the northern sky that have been extensively studied
at a wide range of frequencies. We compared flux densities from the literature with those
derived from AMI SA data in order to examine the calibration of the telescope. The AMI SA flux
densities for each channel are given in Table \ref{tab:oldsnrflux}.
Two SNRs (\SNR(120.1)+(1.4) and
\SNR(130.7)+(3.1)) are structurally extended and the measured AMI flux densities
are corrected for flux loss resulting from changing $uv$ coverage over our
frequency band. The other two SNRs (\SNR(111.7)-(2.1) and \SNR(184.6)-(5.8)) are
less extended and flux loss has a less signficant effect on the AMI flux densities.
We do not give a flux density averaged over the combined bandwidth of AMI SA because of
it is very large. Instead, we provide an interpolated value for the midpoint of AMI's
bandpass (after flux loss correction, where appropriate).

\begin{figure}
\centerline{\includegraphics[width=7.5cm,angle=-90]{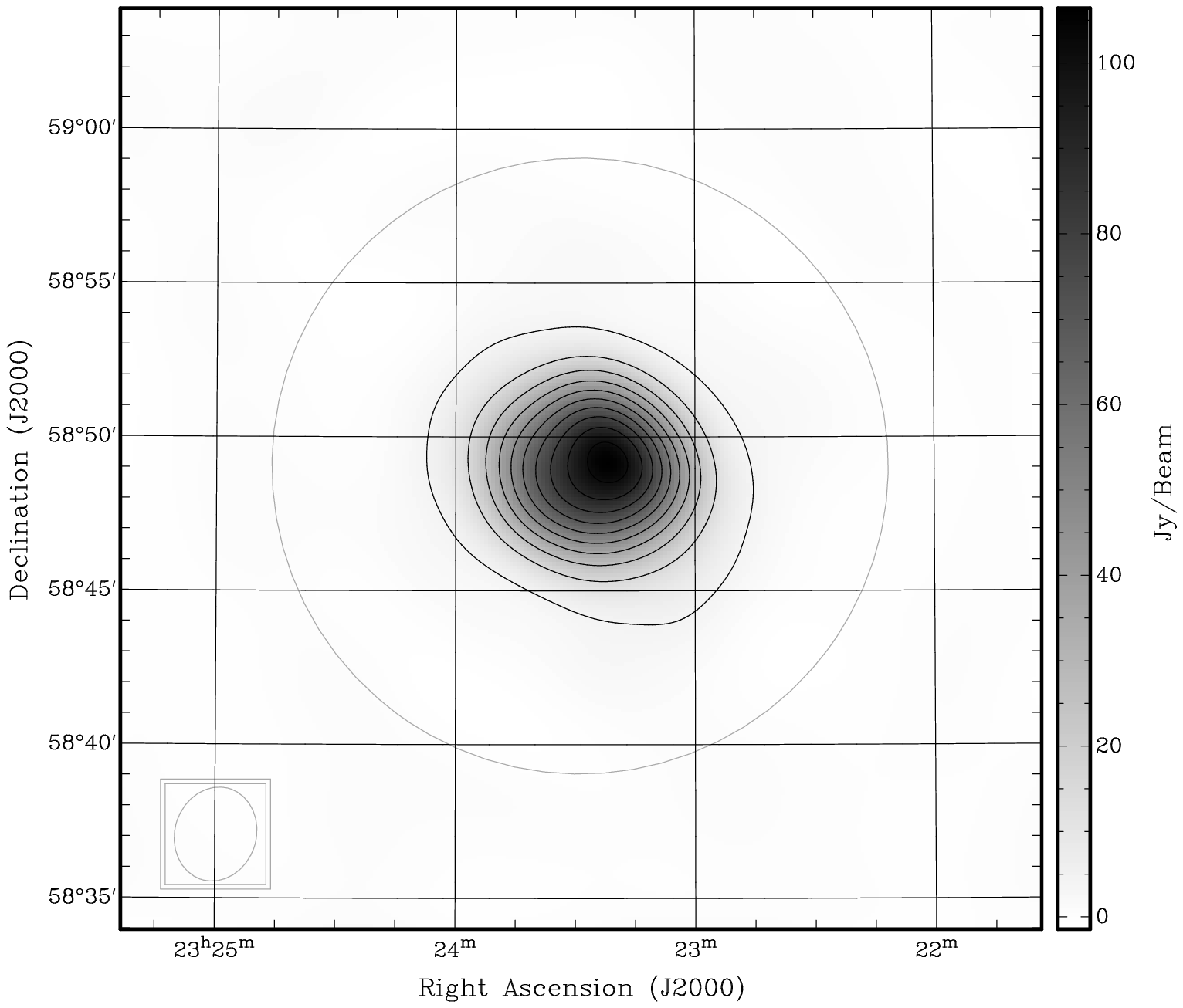}}
\vspace{0.5cm}
\centerline{\includegraphics[width=5cm,angle=-90]{G111_newspec.ps}}
\caption{Above: Map of \SNR(111.7)-(2.1) (Cassiopeia A). AMI SA 16-GHz contours and
greyscale image. Contours are linear percentage levels from 5\% to 95\%
in steps of 10\%. The peak level is 106\,Jy~beam$^{-1}$.
Here, and throughout this paper, the large grey circle indicates the average
FWHM of the primary beam, i.e. where the apparent flux density falls
to 50\%; also, the ellipse in the bottom-left corner indicates the half-maximum
of the synthesized beam; this gives a measure of the resolution of the AMI SA images.
In this, and all images including AMI SA data, the top six channels comprising a bandwidth
 of 5.7\,GHz have been combined; 16\,GHz refers to the central frequency.
Below: Radio spectrum of \SNR(111.7)-(2.1). Data points are from the AMI SA and the dashed
line shows a spectrum fitted to data taken from the literature (including a correction for
secular decay); the hatching indicates the error
on the fit ($\alpha=0.81\pm0.01$).
\label{fig:G111spec}}
\end{figure}

\noindent{\bf \SNR(111.7)-(2.1), Cassiopeia A} (Fig.~\ref{fig:G111spec}).
This is the brightest known Galactic `shell' SNR, and is well studied at all
wavelengths, with its absolute spectrum being the basis of the radio flux
density scale established by \citet{1977A+A....61...99B}. The structure of
Cas~A is barely resolved by the AMI SA observations.

The absolute radio spectrum of
Cas~A follows a power-law, with $\alpha=0.79$ (for epoch 1965.0), from 300\,MHz to
31\,GHz. Since Cas~A is a relatively young SNR, only about 300 years
old (see e.g. \citealt{2002hsr..book.....S}), its flux density shows a significant
secular decrease. \citeauthor{1977A+A....61...99B} constructed their absolute
spectrum for Cas~A by bringing the available absolute flux density measurements
to a common epoch, 1965.0, using a frequency-dependent secular decrease of
\begin{equation}
   \frac{1}{S}
   \frac{{\rm d}S}{{\rm d}t}
      = 0.97(\pm 0.04) - 0.30(\pm 0.04) \log \left( \frac{\nu}{\rm GHz} \right)
        {\rm\ per~cent~yr^{-1}},  \label{eqn:CasA}
\end{equation}
i.e.\ a rate of 0.61~per cent~yr$^{-1}$ at 16\,GHz. From the 1965.0-epoch
spectrum of Cas~A, the predicted flux density at 16\,GHz, epoch 2007.5, is
275$\pm$18\,Jy (assuming an exponential decay at the annual rate given by
eq.~\ref{eqn:CasA}), which is in reasonable agreement with the value obtained with AMI:
256$\pm$25\,Jy. The polarisation of Cas~A is only 1 per cent at 15\,GHz so
this is not expected to affect the measured flux density significantly.

\begin{figure}
\centerline{\includegraphics[width=7.5cm,angle=-90]{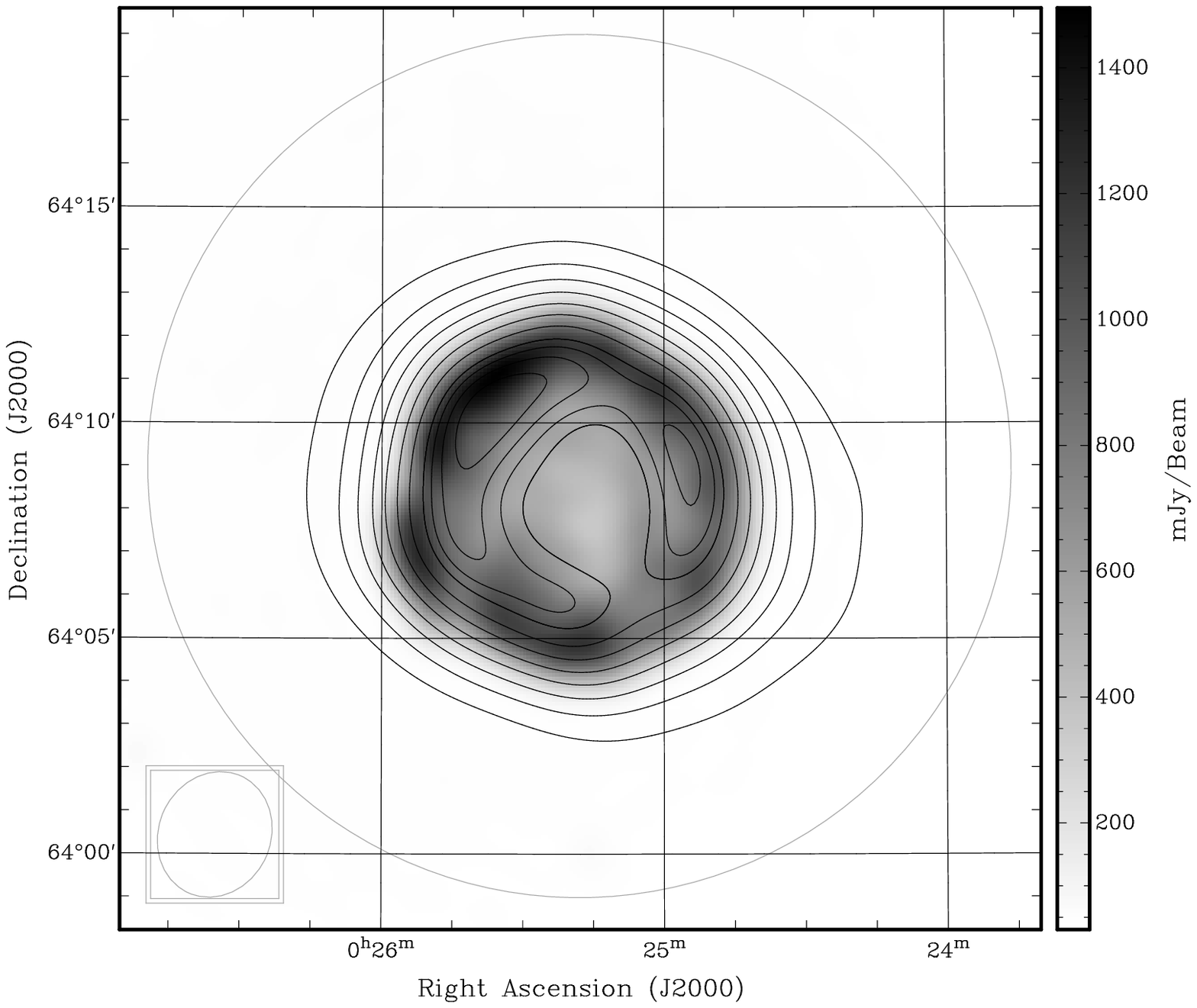}}
\vspace{0.5cm}
\centerline{\includegraphics[width=5cm,angle=-90]{G120_newspec.ps}}
\caption{Above: Map of \SNR(120.1)+(1.4) (Tycho's SNR). AMI SA 16-GHz contours are
overlaid on a CGPS 1.4-GHz greyscale image. Contours are linear percentage
levels from 10\% to 80\% in steps of 10\%, and from 80\% to 100\% in steps of 5\%,
in order to highlight the shell structure.
The peak level is 1.16\,Jy~beam$^{-1}$.
Annotations are as in Fig.~\ref{fig:G111spec}.
Below: Radio spectrum of \SNR(120.1)+(1.4). Data points are from the AMI SA and the dashed
line shows a spectrum fitted to data taken from the literature; the hatching indicates the error
on the fit ($\alpha=0.58\pm0.01$).\label{fig:G120spec}}
\end{figure}

\noindent{\bf \SNR(120.1)+(1.4), Tycho's SNR} (Fig.~\ref{fig:G120spec}).
This is a shell remnant which is clearly resolved by the AMI SA observations. The
shell is brighter to the north-east, in agreement with lower-frequency observations
(e.g. \citealt{2000ApJ...529..453K}). After correction for flux loss,
the flux densities from the AMI SA observations
are in good agreement
with the known radio spectrum of G120.1+1.4 (see Fig.~\ref{fig:G120spec}).
The interpolated value of the flux density at 16\,GHz, 10.8$\pm1.1$\,Jy, is within
the error bars of the predicted value at 16\,GHz, 10.5$\pm0.5$\,Jy.

\begin{figure}
\centerline{\includegraphics[width=7.5cm,angle=-90]{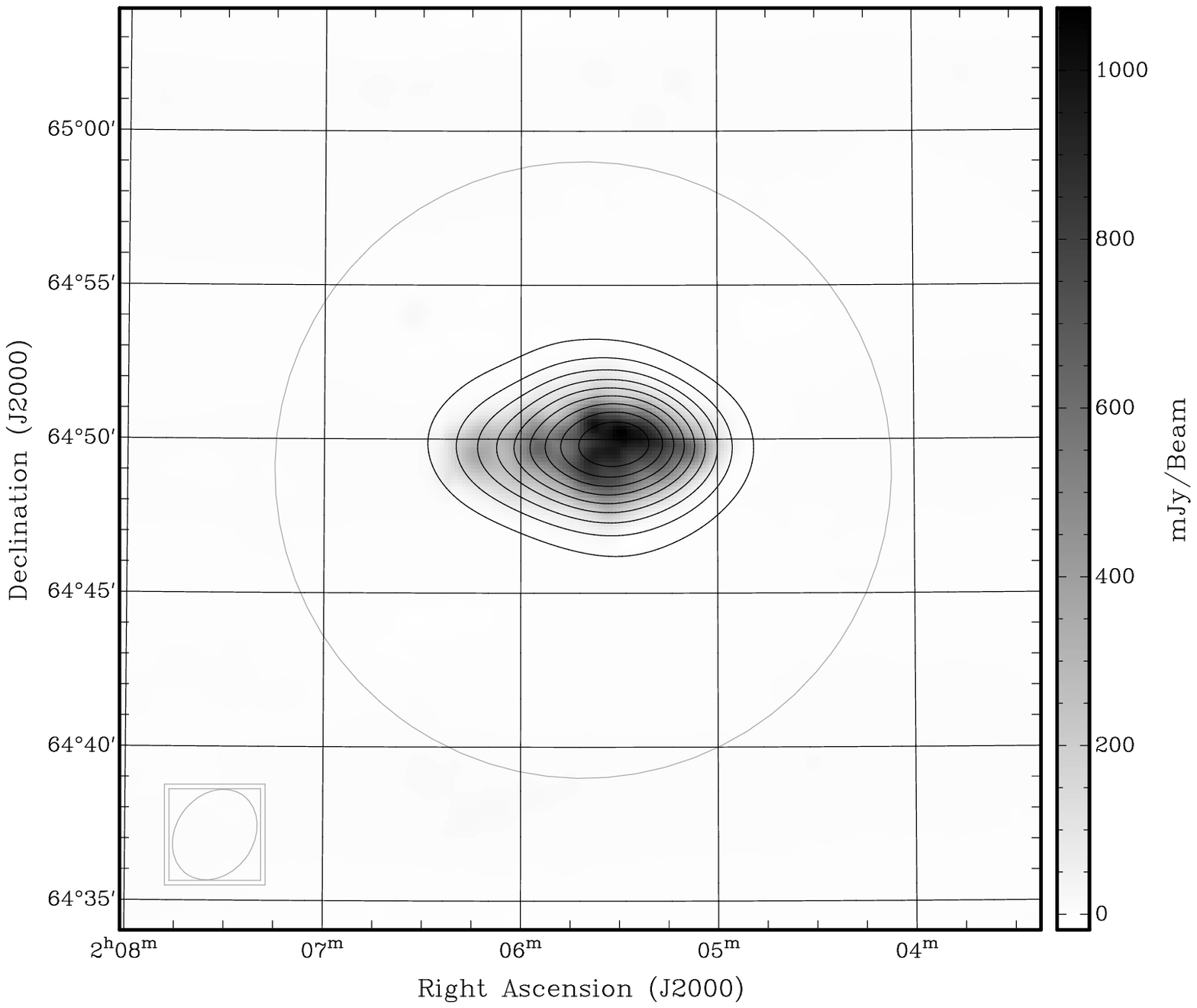}}
\vspace{0.5cm}
\centerline{\includegraphics[width=5cm,angle=-90]{G130_newspec.ps}}
\caption{Above: Map of \SNR(130.7)+(3.1) (3C58). AMI SA 16-GHz contours are
overlaid on a NRAO VLA Sky Survey (NVSS;\citealt{1998AJ....115.1693C})
1.4-GHz greyscale image.
Contours are linear percentage levels from 10\% to 100\% in steps of 10\%.
The peak level is 8.48\,Jy~beam$^{-1}$.
Annotations are as in Fig.~\ref{fig:G111spec}.
Below: Radio spectrum of \SNR(130.7)+(3.1). Data points are from the AMI SA and the dashed
line shows a spectrum fitted to data taken from the literature; the hatching indicates the error
on the fit ($\alpha=0.09\pm0.01$).\label{fig:G130spec}}
\end{figure}

\noindent{\bf \SNR(130.7)+(3.1), 3C58} (Fig.~\ref{fig:G130spec}).
This `filled-centre' SNR may be the remnant of the historical
supernova of {\sc ad}~1181 (see e.g. \citealt{2002hsr..book.....S}), although this association is
not universally accepted (see e.g. \citealt{2008ApJS..174..379F}).
\SNR(130.7)+(3.1) is extended east--west and is known to have a relatively flat radio
spectrum, with $\alpha\approx 0.09$ up to frequencies of at least several tens
of gigahertz (e.g.\ \citealt{1986MNRAS.218..533G, 1989ApJ...338..171S}), but
the position of its spectral turnover is not known \citep{1992MNRAS.258..833G}.
The AMI SA integrated flux densities, after correction for flux loss,
are in reasonable agreement with published flux densities of 3C58
  (e.g. \citealt{1986MNRAS.218..533G}). For instance, our interpolated value
of the flux density at 16\,GHz, 23.1$\pm2.3$\,Jy, agrees well with the predicted value
 at 16\,GHz, 25.0$\pm1.1$\,Jy.

\begin{figure}
\centerline{\includegraphics[width=7.5cm,angle=270]{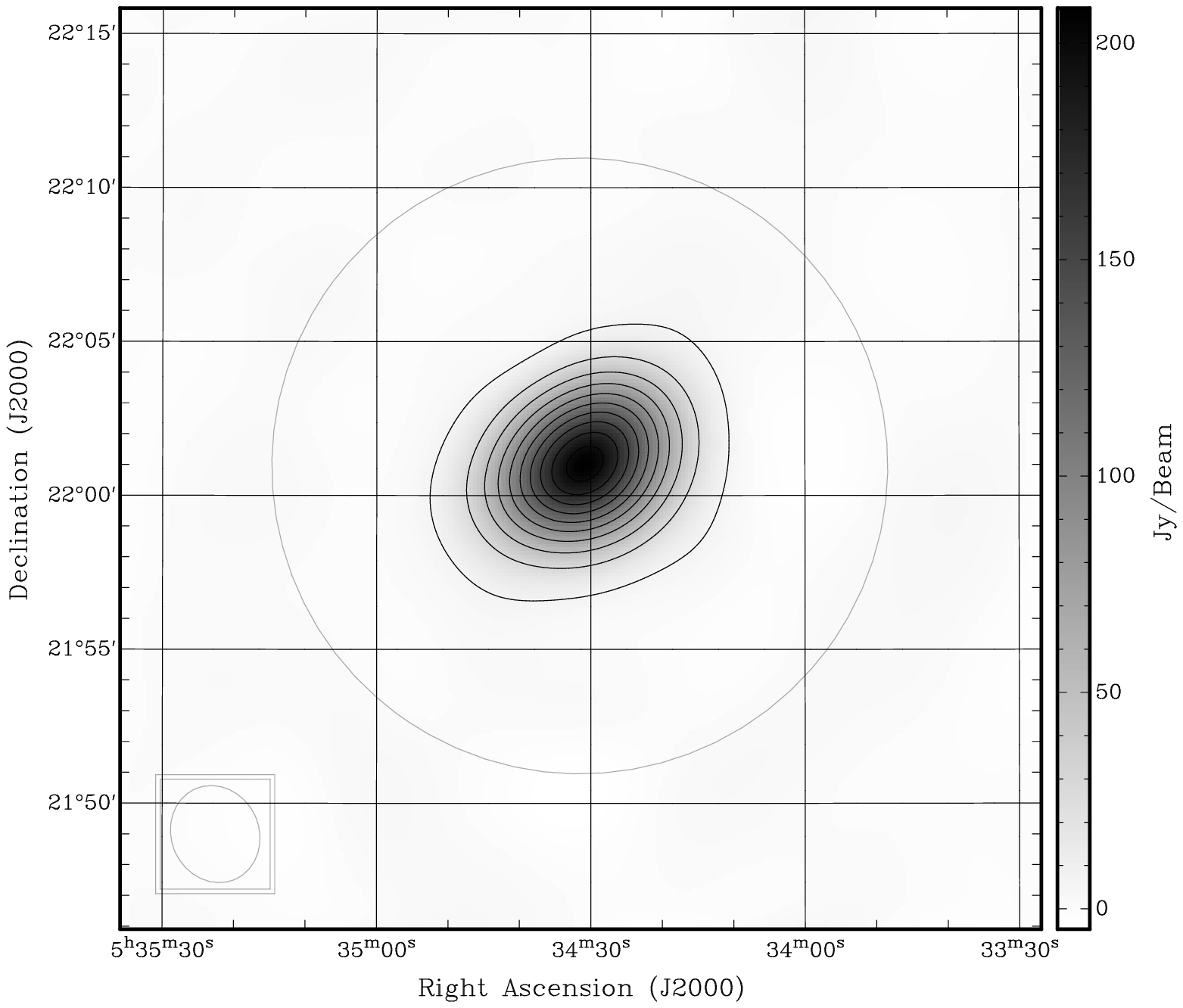}}
\vspace{0.5cm}
\centerline{\includegraphics[width=5cm,angle=-90]{G184_newspec.ps}}
\caption{
Above: Map of \SNR(184.6)-(5.8) (Crab Nebula). AMI SA 16-GHz contour
and greyscale image.  Contours and annotations are as in
Fig.~\ref{fig:G130spec}.  The peak level is 208\,Jy~beam$^{-1}$.
Below: Radio spectrum of \SNR(184.6)-(5.8). Data points are from the AMI SA and the dashed
line shows a spectrum fitted to data taken from the literature (including a correction for
secular decay); the hatching indicates the error
on the fit ($\alpha=0.294\pm0.002$).\label{fig:G184spec}}
\end{figure}

\noindent{\bf \SNR(184.6)-(5.8), the Crab Nebula} (Fig.~\ref{fig:G184spec}).
The spectrum of the Crab Nebula is well known, showing a power-law behaviour
over a wide range of radio frequencies (e.g.\ \citealt{1977A+A....61...99B})
with $\alpha=0.30$, with a secular decrease of $\approx 0.17$~per~cent~yr$^{-1}$
(\citealt{1985ApJ...293L..73A}, see also \citealt{2007ARep...51..570V}).
Recent observations by \citet{2004MNRAS.355.1315G}
also show that the Crab Nebula has a very similar structure
from 1.4 to 345\,GHz, consistent with a simple, single
synchroton spectral power-law between these frequencies. We note that the
Crab Nebula has $\approx 9$~per cent integrated polarization at 15\,GHz
\citep{1966ApJ...144..437B}, at a position angle of $140^{\circ}$. For these
integrated polarization parameters, the observed I~+~Q flux density will
overestimate the true flux density by about 3.5~per~cent, so a correction
of this magnitude has been
made to the AMI flux densities listed in Table~\ref{tab:oldsnrflux}.

We assume a decrease in flux density of 7~per~cent over
the 40 years since the absolute flux
density measurements used for \citeauthor{1977A+A....61...99B}'s spectrum,
based on a decrease of $\sim 0.17$~per~cent~yr$^{-1}$.
Taking this into account, the integrated flux densities from the AMI SA observations are in good agreement
with the results expected from extrapolation of the results of
\citeauthor{1977A+A....61...99B}. For comparison, an interpolated value of
the measured flux density at 16\,GHz is 450$\pm$45\,Jy, compared to an predicted
value of 424$\pm$33\,Jy. This object has not been flux-loss corrected but
as it is relatively compact, this is not expected to be a large effect.

\begin{table*}
\caption{AMI SA I+Q flux densities of well-studied SNR. Above: corrected for flux loss, where possible; below: original flux densities for those sources which undergo a flux loss correction.\label{tab:oldsnrflux}}
\begin{tabular}{lcccccc}\hline
 & \multicolumn{6}{c}{Freq. /GHz}\\ \cline{2-7}
 & 14.2 & 15.0 & 15.7 & 16.4 & 17.1 & 17.9\\
 Name  & (Jy) & (Jy) & (Jy) & (Jy) & (Jy) & (Jy)\\\hline
 \SNR(111.7)-(2.1) $^{a}$ & $290\pm29$ & $269\pm27$ & $263\pm26$ & $233\pm23$ & $228\pm23$ & $216\pm22$  \\
 \SNR(120.1)+(1.4) $^{b}$ & -- & $11.2\pm0.6$ & $11.0\pm0.6$ & $10.4\pm0.5$ & $10.4\pm0.5$ & --  \\
 \SNR(130.7)+(3.1) $^{b}$ & $25.8\pm2.4$ & $25.0\pm2.4$ & $23.9\pm2.3$ & $23.2\pm2.1$ & $23.2\pm2.1$ & $23.8\pm1.2$ \\
 \SNR(184.6)-(5.8) $^{a}$ & $452\pm45$ & $463\pm46$ & $462\pm46$ & $447\pm45$ & $433\pm43$ & $435\pm44$ \\ \hline
 \SNR(120.1)+(1.4) & -- & $10.3\pm0.3$ & $9.8\pm0.3$ & $8.8\pm0.2$ & $7.7\pm0.2$ & --  \\
 \SNR(130.7)+(3.1) & $24.7\pm1.2$ & $24.3\pm1.2$ & $23.4\pm1.1$ & $21.8\pm1.0$ & $21.2\pm1.0$ & $22.4\pm1.1$ \\ \hline
\end{tabular}
\begin{minipage}{16cm}{
 $^{a}$Uncorrected for flux loss; no CGPS coverage.

 $^{b}$Corrected for flux loss using CGPS data.
}
\end{minipage}
\end{table*}
\subsection{Other SNRs at 16\,GHz}
These SNRs are less well-studied and most have not been observed at frequencies greater than 10\,GHz
so high frequency data from AMI SA are of particular interest.
We present short notes on each SNR and derive spectral indices (shown in the captions to each image) from a
combination of AMI SA data and flux densities from the literature at lower frequencies,
using the method described in Section~\ref{sec:data}. Some exceptions are made:
\SNR(59.5)+(0.1) has too complicated a structure to reliably fit to, and
\SNR(74.9)+(1.2) has a spectral break at 11\,GHz, so the fit is only made to
data below this frequency.

\begin{figure}
\centerline{\includegraphics[width=7.5cm,keepaspectratio,angle=-90]{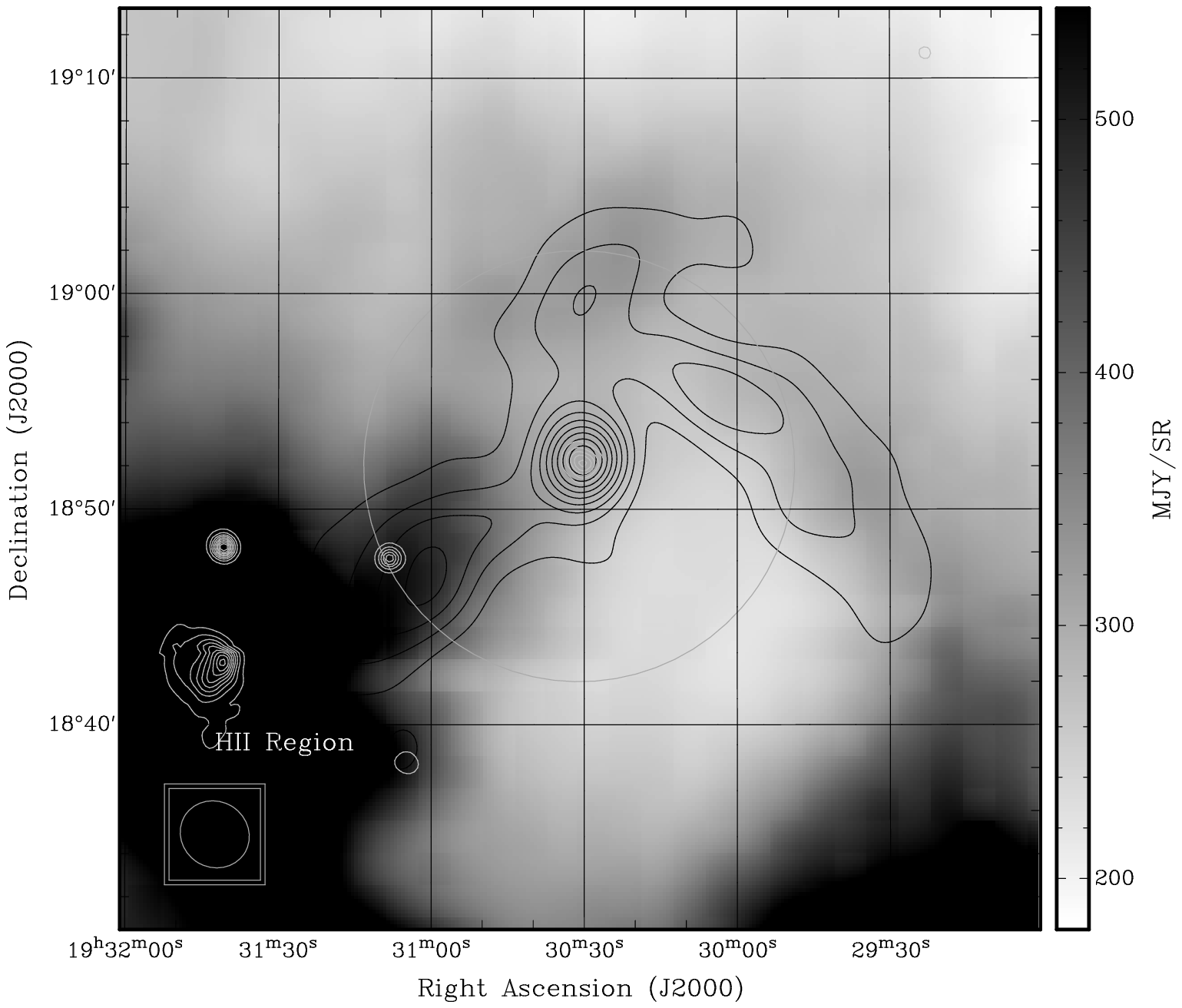}}
\vspace{0.5cm}
\centerline{\includegraphics[width=5cm,angle=-90]{G54_newspec.ps}}
\caption{Above: Map of \SNR(54.1)+(0.3). AMI SA 16-GHz contours (black) and
NVSS 1.4-GHz contours (grey) are overlaid on a IRAS 100-$\mu$m greyscale
image. The peak level of the AMI SA contours is 0.21\,mJy~beam$^{-1}$ and the
peak level of the NVSS contours is 389\,mJy~beam$^{-1}$.
Contours and annotations are as in Fig.~\ref{fig:G130spec}.
Below: Radio spectrum of \SNR(54.1)+(0.3).
Integrated flux densities taken from the literature (Table~\ref{tab:G54flux})
are shown as crosses, and those from AMI SA (Table~\ref{tab:newsnrflux})
are shown as filled circles.
The best-fitting power-law, with $\alpha=0.14$, is shown as a dashed line. \label{fig:G54spec}}
\end{figure}

\noindent{\bf \SNR(54.1)+(0.3)} (Fig.~\ref{fig:G54spec}).
This is a small filled-centre remnant for which there are no published radio
flux densities at frequencies above 5\,GHz, so the position of any possible spectral break is
not yet known. We derive an integrated flux density at 10\,GHz from the data of
\citet{1987PASJ...39..709H}.
Three `arms' of extended emission surround the remnant and are
obvious in both the 2.7-GHz data of
the Effelsberg 11-cm survey \citep{1984A+AS...58..197R,1990A+AS...85..805F}
and in infra-red emission seen in the IRAS 100-$\mu$m image. To the south-east
a large {\sc Hii} region (e.g. \citet{1987A+A...181..378C}) can be
seen which, like the `arms', is well-traced by the IRAS 100\,$\mu$m
data. This background morphology makes fitting the flux density of
\SNR(54.1)+(0.3) difficult as can be seen by the scatter in the flux density data from
the literature -- see Table~\ref{tab:G54flux}. The AMI SA observations
indicate a steeper spectrum than at lower frequencies, however the
reliability of this result is brought into question by the varying background
on different angular scales which may affect the flux densities and lead to an
artificially steep spectrum. In an attempt to assess this effect the data from
each AMI SA channel were truncated to a common $uv$ range. This produced a
similar spectrum across the AMI SA band, suggesting that the steepening is
indeed genuine. However, in order to provide a definite
answer observations at higher frequencies are required.

\begin{table}
\centering
\caption{Integrated flux densities from the literature for
\SNR(54.1)+(0.3).\label{tab:G54flux}}
\begin{tabular}{ccc} \hline
 $\nu$/GHz & $S_{\rm{i}}$/mJy & Reference \\ \hline
 0.327 & $495\pm75$ & \citet{1988AJ.....95.1162V} \\
 0.327 & $504\pm17$ & \citet{1996ApJS..107..239T} \\
 1.4   & $478\pm30$ & \citet{1988AJ.....95.1162V} \\
 1.4   & $327.5\pm10.9$ & \citet{1989AJ.....97.1064C} \\
 1.42  & $364\pm36$ & \citet{1987A+A...171..261C} \\
 1.6   & $417\pm30$ & \citet{1988AJ.....95.1162V} \\
 2.7   & $580\pm60$ & \citet{1984A+AS...58..197R} \\
 4.75  & $370\pm40$ & \citet{1985A+A...151L..10R} \\
 4.8   & $325\pm20$ & \citet{1988AJ.....95.1162V} \\
 4.875 & $400\pm40$ & \citet{1979A+AS...35...23A} \\
 5.0   & $306\pm31$ & \citet{1990ApJS...74..129G} \\
 10.7  & $263\pm30$ & This work, from the \\
       &            & data of \citet{1987PASJ...39..709H} \\ \hline
\end{tabular}
\end{table}

\begin{figure}
\centerline{\includegraphics[width=7.5cm,angle=-90]{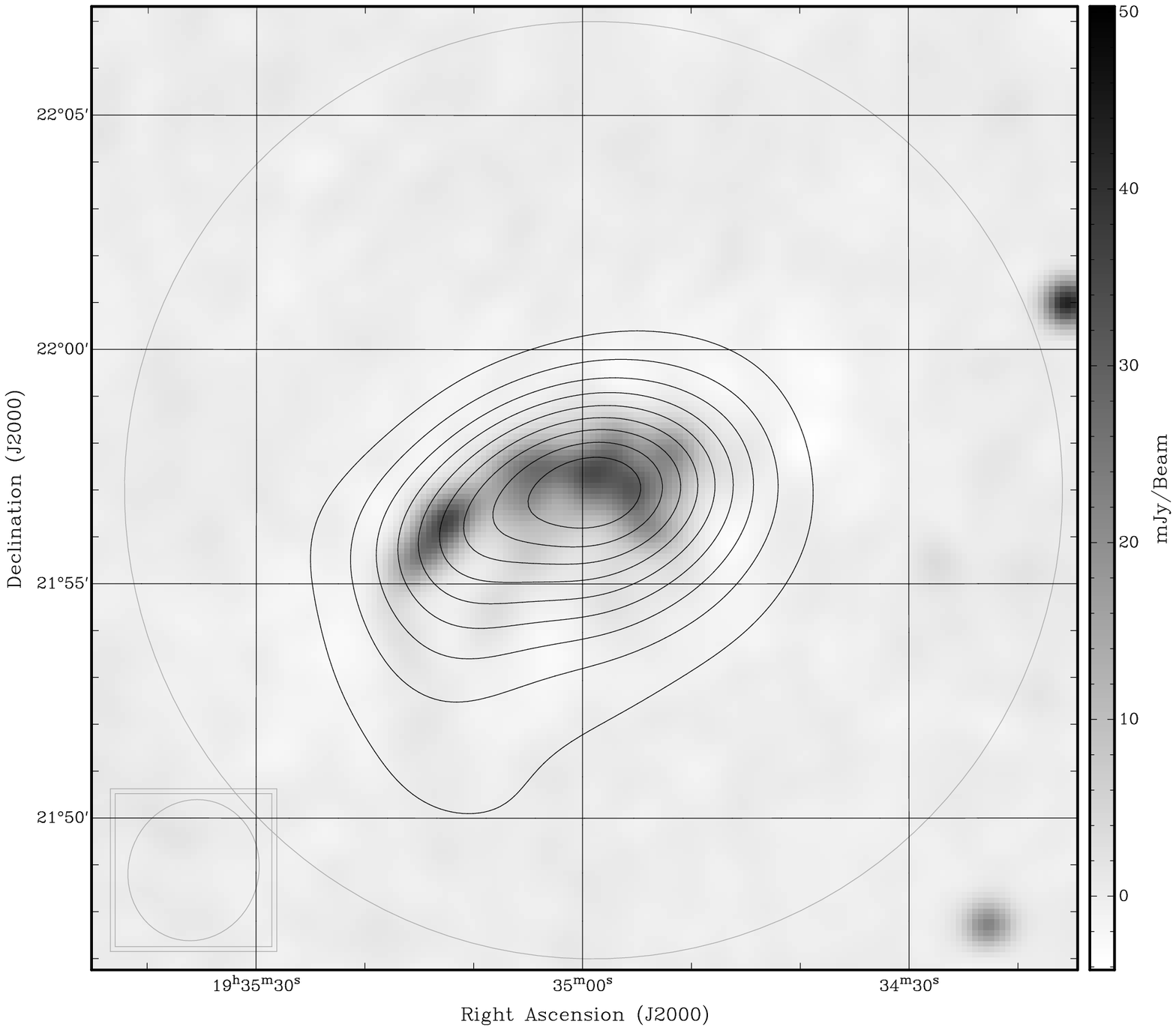}}
\vspace{0.5cm}
\centerline{\includegraphics[width=5cm,angle=-90]{G57_newspec.ps}}
\caption{Above: Map of the \SNR(57.2)+(0.8) region. AMI SA 16-GHz contours are
overlaid on a NVSS 1.4-GHz greyscale image. Contours and annotations are as in Fig.\ref{fig:G130spec}.
The peak level is 92.0\,mJy~beam$^{-1}$.
Below: Radio spectrum of \SNR(57.2)+(0.8).
Integrated flux densities taken from the literature (Table~\ref{tab:G57flux})
are shown as crosses, and those from AMI SA (Table~\ref{tab:newsnrflux})
are shown as filled circles.
The best-fitting power-law, with $\alpha=0.67$, is shown as a dashed line.\label{fig:G57spec}}
\end{figure}

\noindent{\bf \SNR(57.2)+(0.8)} (Fig.~\ref{fig:G57spec}).
This remnant has not previously been studied at
frequencies above 5\,GHz. In the AMI SA band it appears to have a teardrop-shaped emission limb of peak flux
0.09\,Jy~beam$^{-1}$ at 16\,GHz. This SNR is not associated with
PSR1937+214, which is more than 1 degree away. The AMI SA flux densities have not been
flux corrected as there is no CGPS coverage at this position. This may explain the observed
slightly lower fluxes and the apparent
steepening of the spectrum across the AMI band compared to the extrapolated spectrum
from low frequencies.

\begin{table}
\centering
\caption{Integrated flux densities from the literature for
\SNR(57.2)+(0.8).\label{tab:G57flux}}
\begin{tabular}{ccc} \hline
 $\nu$/GHz & $S_{\rm{i}}$/Jy & Reference \\ \hline
 0.083 & $8.0\pm2.0$ & \citet{1994ARep...38...95K} \\
 0.178 & $4.0\pm1.0$ & \citet{1987A+A...171..261C} \\
 0.327 & $3.2\pm0.1$ & \citet{1996ApJS..107..239T} \\
 1.4   & $1.29\pm0.13$ & \citet{1992ApJS...79..331W} \\
 1.42  & $1.43\pm0.14$ & \citet{1990A+AS...85..805F} \\
 1.42  & $1.34\pm0.10$ & \citet{1984A+A...130..257S} \\
 2.7   & $0.90\pm0.09$ & \citet{1984A+AS...58..197R} \\
 2.7   & $0.86\pm0.10$ & \citet{1984A+A...130..257S} \\
 3.9   & $0.8\pm0.1$ & \citet{1987AISAO..25...84T} \\
 4.85  & $0.618\pm0.062$ & \citet{1991ApJS...75....1B} \\
 4.85  & $0.642\pm0.023$ & \citet{1996ApJS..103..427G} \\
 4.85  & $0.64\pm0.02$ & \citet{1996ApJS..107..239T} \\ \hline
\end{tabular}
\end{table}

\begin{figure}
\centerline{\includegraphics[width=7.5cm,angle=-90]{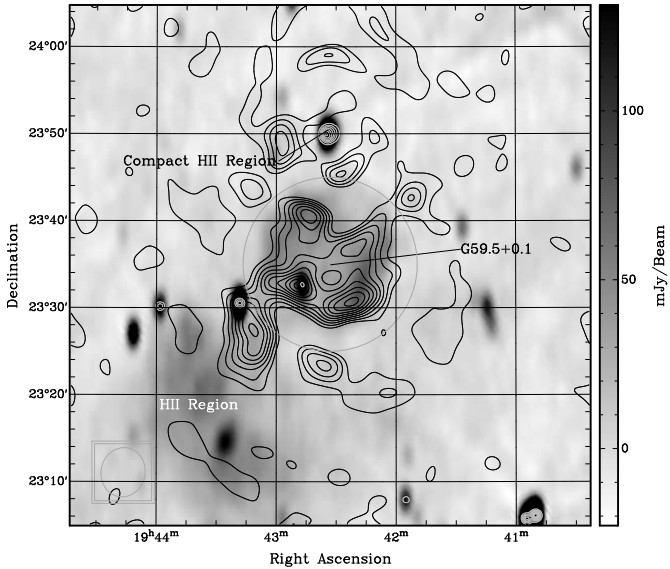}}
\caption{Above: Map of \SNR(59.5)+(0.1). AMI SA 16-GHz contours (black) and
NVSS 1.4-GHz contours (grey) are overlaid on a WSRT 327-MHz greyscale image. The peak
level of the AMI SA contours is 18.8\,mJy~beam$^{-1}$ and the peak level of the
NVSS contours is 417\,mJy~beam$^{-1}$.
The north compact {\sc Hii} region has been subtracted from the AMI SA data to enhance
the image of \SNR(59.5)+(0.1).
Contours and annotations are as in Fig.~\ref{fig:G130spec}.} \label{fig:G59spec}
\end{figure}

\noindent{\bf \SNR(59.5)+(0.1)} (Fig.~\ref{fig:G59spec}).
The AMI SA observation of \SNR(59.5)+(0.1) is dominated by a bright {\sc Hii}
region (e.g. \citealt{1989ApJS...71..469L}) which lies to the north of the SNR. In order to
produce a better image of the SNR, the {\sc Hii} region was observed separately. The {\sc Hii} region
 was then \textsc{clean}ed and the \textsc{clean} components (scaled for the primary beam)
subtracted from the original \SNR(59.5)+(0.1) visibility data.
These data were then re-mapped and \textsc{clean}ed using NVSS and
Westerbork Synthesis Radio Telescope (WSRT) 327-MHz data \citep{1992AJ....103..931T} as a reference, to produce the image
shown in Fig.~\ref{fig:G59spec}. As the AMI SA's beam reaches a null at 40\arcmin  from the pointing centre, the {\sc Hii} region S86 \citep{1989ApJS...71..469L} is revealed only as a small limb to the southeast. We confirm that
 \SNR(59.5)+(0.1) is a shell-type SNR with the `barrel-like' morphology described by \citet{1996ApJS..107..239T}: brighter in
the SW and NE. The flux in the SE is contaminated with an extragalactic source
as detected by NVSS. The slight NW extension is also influenced by a point
source. It is difficult to image north of $\delta =
23^{\circ}45$\arcmin as the subtraction of the {\sc Hii} region is imperfect and
may have left artefacts.

\citet{1996ApJS..107..239T} measure the flux density of \SNR(59.5)+(0.1) to be
$5.07\pm0.16$\,Jy at 327\,MHz and $< 1.65$\,Jy at 4.85\,GHz, suggesting a
spectral index of greater than $\alpha = 0.41$. This would imply a flux density
of $\leq 1$\,Jy at 16\,GHz. From the AMI SA combined channel map, see
Fig.~\ref{fig:G59spec}, we measure a flux density of $S_{\rm{i}} =
0.146\pm0.015$\,Jy, where the error has been estimated as $\sigma =
\sqrt{\sigma_{\rm{th}}^2+(0.1S_{\rm{i}})^2}$. A 10 per cent error on the flux density
is used here since the integrated flux density is highly dependent on the defined
fitting region. Although the SNR catalogue of Green (2004) lists the angular
size of \SNR(59.5)+(0.1) to be 5\arcmin, this is an error, as
\citet{1996ApJS..107..239T} measure the extent of this source to be 15\arcmin,
a size which is consistent with the emission seen at 16\,GHz by AMI SA.
We do not list the AMI SA flux densities for individual channels as the
complex nature of the region makes flux loss estimates unreliable. This
prevents us from fitting a spectrum to the measured flux densities since for a
source of such angular extent and distinct substructure the losses
will not only be large, but will also differ significantly between
channels.

\begin{figure}
\centerline{\includegraphics[width=7.5cm,angle=-90]{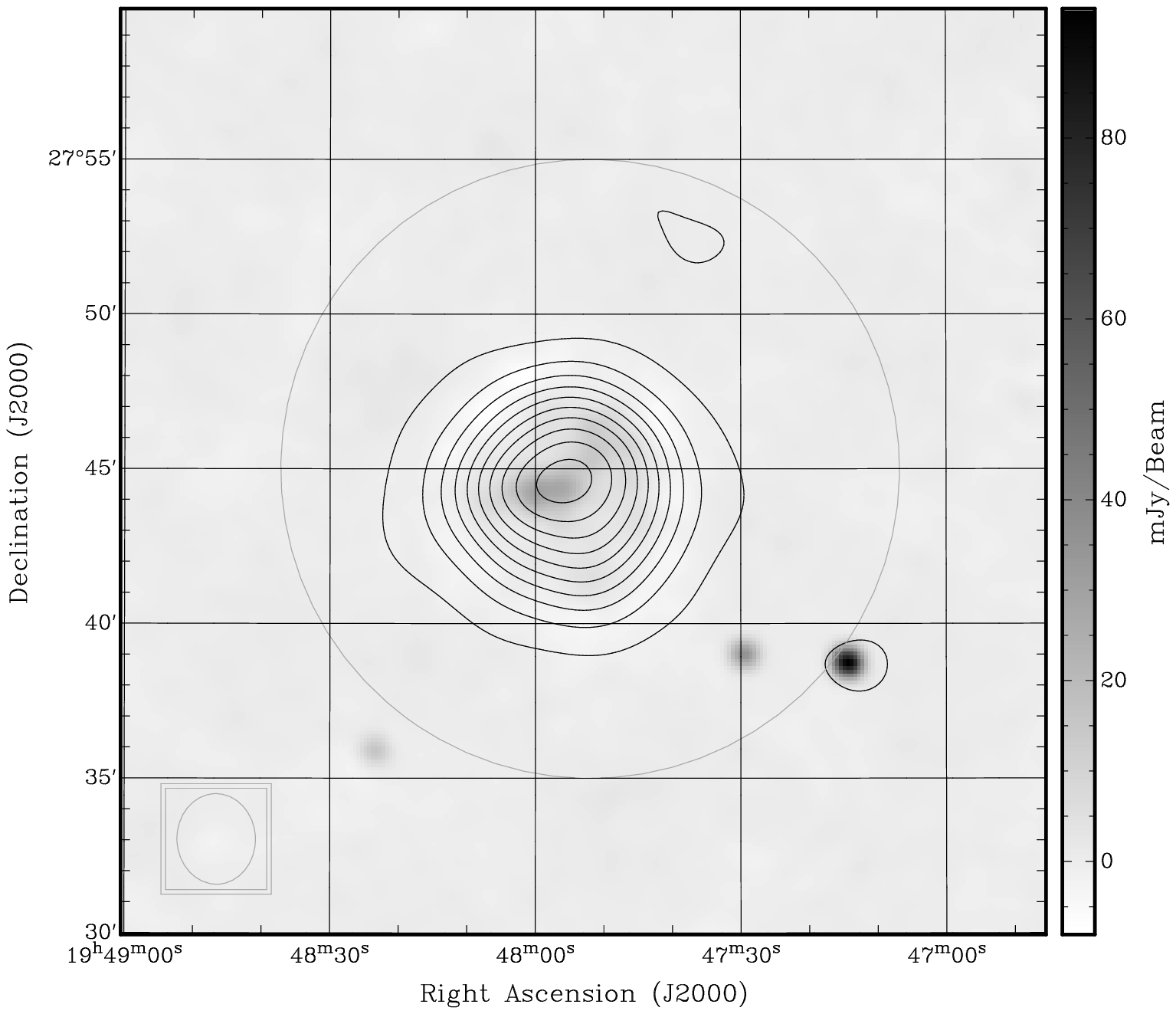}}
\vspace{0.5cm}
\centerline{\includegraphics[width=5cm,angle=-90]{G63_newspec.ps}}
\caption{Above: Map of \SNR(63.7)+(1.1). AMI SA 16-GHz contours are overlaid on
a NVSS 1.4-GHz greyscale image. Contours and annotations are as in
Fig.~\ref{fig:G111spec}. The peak level is 153\,mJy~beam$^{-1}$.
Below: Radio spectrum of \SNR(63.7)+(1.1).
Integrated flux densities taken from the literature (Table~\ref{tab:G63flux})
are shown as crosses, and those from AMI SA (Table~\ref{tab:newsnrflux})
are shown as filled circles.
The best-fitting power-law, with $\alpha=0.30$, is shown as a dashed line. \label{fig:G63spec}}
\end{figure}

\noindent{\bf \SNR(63.7)+(1.1)} (Fig.~\ref{fig:G63spec}).
The AMI SA observation shows this to be a relatively featureless object.
The integrated flux densities listed here, see
Table~\ref{tab:G63flux}, include the flux from two nearby compact sources which
are unresolved in AMI SA and earlier observations \citep{1997AJ....114.2068W}.
The AMI SA flux densities are consistent with the spectrum extrapolated from lower
frequencies.

\begin{table}
\centering
\caption{Integrated flux densities from the literature for \SNR(63.7)+(1.1).\label{tab:G63flux}}
\begin{tabular}{ccc}\hline
 $\nu$/GHz & $S_{\rm{i}}$/Jy & Reference \\\hline
 0.327 & $2.47\pm0.25$ & \citet{1996ApJS..107..239T}\\
 0.408 & $2.1\pm0.2$ & \citet{1997AJ....114.2068W} \\
 1.39 & $1.76\pm0.01$ & \citet{1997AJ....114.2068W} \\
 1.42 & $1.71\pm0.02$ & \citet{1997AJ....114.2068W} \\
 1.42 & $1.62\pm0.16$ & \citet{1990A+AS...85..805F} \\
 1.42 & $1.34\pm0.1$ & \citet{1984A+A...130..257S}  \\
 2.695 & $1.5\pm0.15$ & \citet{1984A+AS...58..197R} \\
 2.695 & $1.41\pm0.1$ & \citet{1997AJ....114.2068W} \\
 4.85 & $1.147\pm0.115$ & \citet{1991ApJS...75....1B} \\
 4.85 & $1.381\pm0.138$ & \citet{1992ApJS...79..331W} \\
 4.85 & $1.16\pm0.1$ & \citet{1997AJ....114.2068W} \\
 10.55 & $0.95\pm0.05$ & \citet{1997AJ....114.2068W} \\\hline
\end{tabular}
\end{table}

\begin{figure}
\centerline{\includegraphics[width=7.5cm,angle=-90]{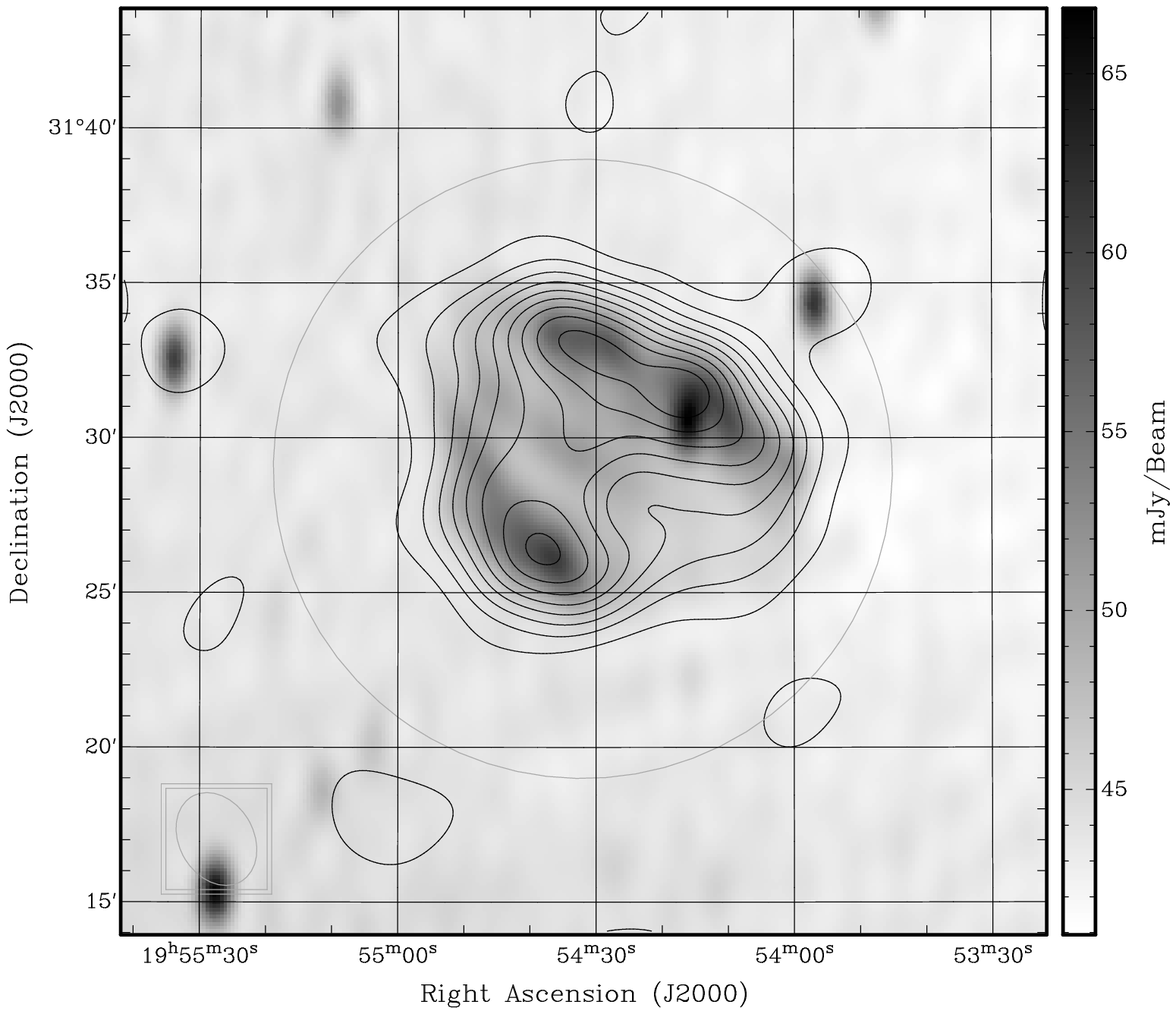}}
\vspace{0.5cm}
\centerline{\includegraphics[width=5cm,angle=-90]{G67_newspec.ps}}
\caption{Above: Map of \SNR(67.7)+(1.8). AMI SA 16-GHz contours are overlaid on
a CGPS 1.4-GHz greyscale image. Contours and annotations are as in Fig.~\ref{fig:G130spec}.
The peak level is 17.3\,mJy~beam$^{-1}$.
Below: Radio spectrum of \SNR(67.7)+(1.8).
Integrated flux densities taken from the literature (Table~\ref{tab:G67flux})
are shown as crosses, and those from AMI SA (Table~\ref{tab:newsnrflux})
are shown as filled circles.
The best-fitting power-law, with $\alpha=0.62$, is shown as a dashed line.\label{fig:G67spec}}
\end{figure}

\noindent{\bf \SNR(67.7)+(1.8)} (Fig.~\ref{fig:G67spec}).
This remnant has a double arc structure, confused with an extragalactic point
source near $19^{\rm h} 54^{\rm m} 15^{\rm s}$, $+31^{\circ} 30\arcmin 45\arcsec$.
The AMI SA flux densities are in good agreement with extrapolation from the available
lower frequency flux densities.
\begin{table}
\centering
\caption{Integrated flux densities from the literature for
\SNR(67.7)+(1.8).\label{tab:G67flux}}
\begin{tabular}{ccc}\hline
  $\nu$/GHz & $S_{\rm{i}}$/Jy & Reference \\ \hline
  0.327 & $1.88\pm0.08$ & \citet{1992AJ....103..931T} \\
  1.42 & $0.72\pm0.07$ & \citet{1990A+AS...85..805F} \\
  3.90 & $0.60\pm0.10$ & \citet{1996BSAO...41...64T} \\
  4.85 & $0.42\pm0.05$ & \citet{1992AJ....103..931T} \\ \hline
\end{tabular}
\end{table}

\begin{figure}
\centerline{\includegraphics[width=7.5cm,angle=-90]{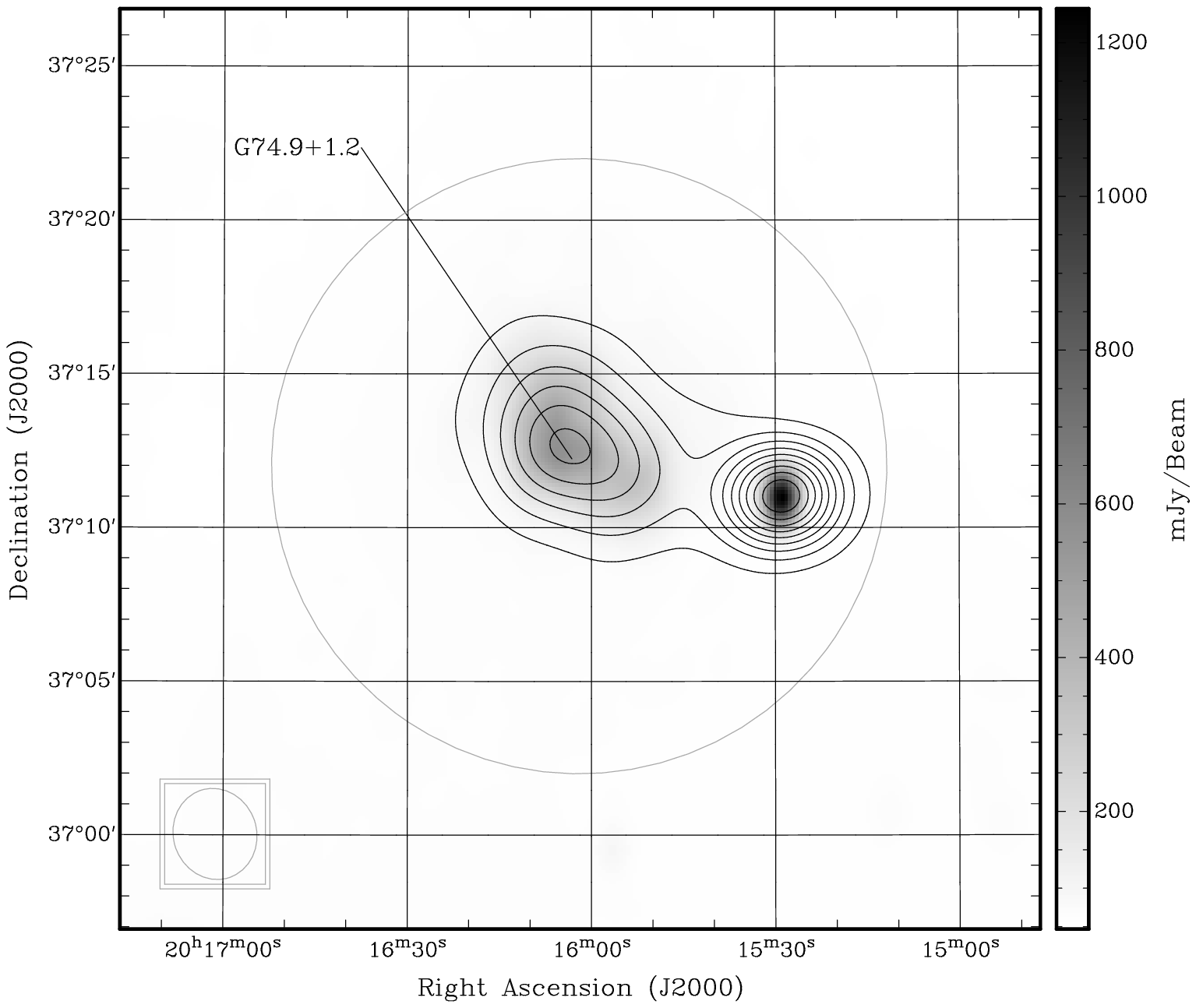}}
\vspace{0.5cm}
\centerline{\includegraphics[width=5cm,angle=-90]{G74_newspec.ps}}
\caption{Above: Map of \SNR(74.9)+(1.2). AMI SA 16-GHz contours are overlaid on
a CGPS 1.4-GHz greyscale image. Contours and annotations are as in Fig.~\ref{fig:G130spec}.
The peak level is 1.38\,Jy~beam$^{-1}$.
Below: Radio spectrum of \SNR(74.9)+(1.2).
Integrated flux densities taken from the literature (Table~\ref{tab:G74flux})
are shown as crosses, and those from AMI SA (Table~\ref{tab:newsnrflux})
are shown as filled circles.
The best-fitting power-law to the data below 12\,GHz, with $\alpha=0.26$, is shown as a dashed line.\label{fig:G74spec}}
\end{figure}

\noindent{\bf \SNR(74.9)+(1.2)} (Fig.~\ref{fig:G74spec}).
This is a filled-centre remnant, with a nearby flat-spectrum extraglactic
source. At lower radio frequencies this remnant has a relatively flat spectrum,
with $\alpha=0.26$, with observations from \citet{1987A+AS...71..189M} and \citet{1989ApJ...338..171S}
implying a spectral break at around 11\,GHz, i.e.\ at a
frequency just below the AMI SA band. The AMI SA flux densities are indeed less than
the expected values from a simple extrapolation from lower frequencies, in agreement
with the 14.35-GHz flux density from \citet{2000AJ....119.2801L}.

\begin{table}
\centering
\caption{Integrated flux densities from the literature for
\SNR(74.9)+(1.2).\label{tab:G74flux}}
\begin{tabular}{ccc}\hline
 $\nu$/GHz & $S_{\rm{i}}$/Jy & Reference \\\hline
 0.408 & $ 11.6\pm  0.4$ & \citet{1991A+A...241..551W} \\
 0.408 & $ 13.3\pm  0.8$ & \citet{1990MNRAS.246..169P} \\
 0.61 & $ 9.1\pm   1.2$ & \citet{1978A+A....70..389W} \\
 1.42 & $7.2\pm   0.3$ & \citet{1990MNRAS.246..169P} \\
 1.415 & $8.0\pm   1.5$ & \citet{1975A+A....38..461D} \\
 1.415 & $8.7 \pm  1.2$ & \citet{1978A+A....70..389W} \\
 2.695 & $7.6\pm   0.5$ & \citet{1980A+A....84..237G} \\
 4.80 & $7.5\pm   0.7$ & \citet{1991A+A...241..551W} \\
 4.995 & $5.6\pm   1.3$ & \citet{1978A+A....70..389W} \\
 8.35 & $4.89\pm  0.50$ & \citet{2000AJ....119.2801L} \\
 10.69 & $4.8 \pm  1.2$ & \citet{1975A+A....44..187G} \\
 14.35 & $2.47 \pm 0.20$ & \citet{2000AJ....119.2801L} \\
 32.0 & $1.47\pm  0.19$ & \citet{1987A+AS...71..189M} \\
 84.2 & $ 0.6\pm 0.1$ & \citet{1989ApJ...338..171S} \\\hline
\end{tabular}
\end{table}

\begin{figure}
\centerline{\includegraphics[width=7.5cm,angle=-90]{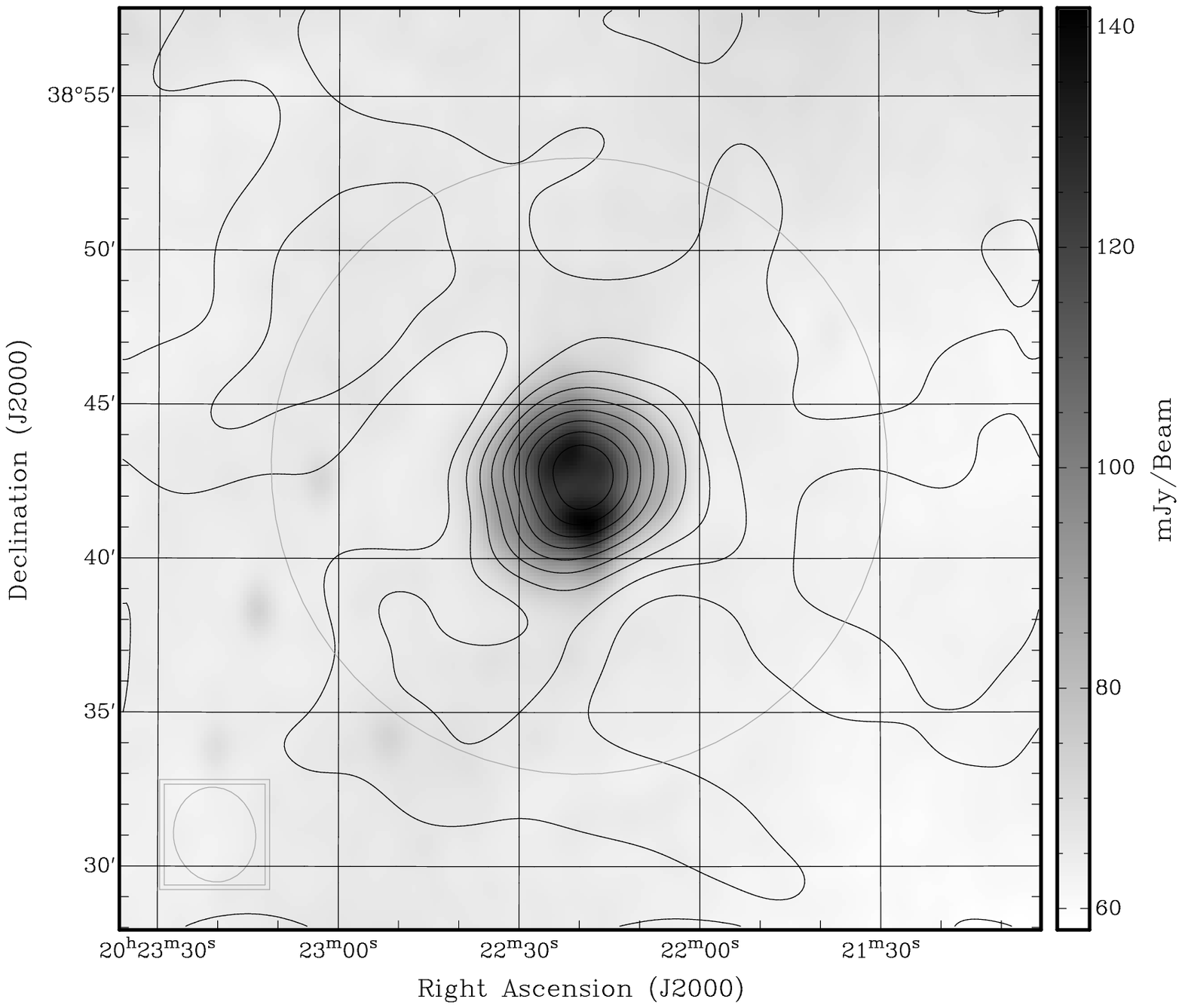}}
\vspace{0.5cm}
\centerline{\includegraphics[width=5cm,angle=-90]{G76_newspec.ps}}
\caption{Above: Map of \SNR(76.9)+(1.0). AMI SA 16-GHz contours are overlaid on
a CGPS 1.4-GHz greyscale image.  Contours and annotations are as in Fig.~\ref{fig:G130spec}.
The peak level is 46.6\,mJy~beam$^{-1}$.
Below: Radio spectrum of \SNR(76.9)+(1.0).
Integrated flux densities taken from the literature (Table~\ref{tab:G76flux})
are shown as crosses, and those from AMI SA (Table~\ref{tab:newsnrflux})
are shown as filled circles.
The best-fitting power-law to the data above 1\,GHz, with $\alpha=0.75$, is shown as a dashed line.\label{fig:G76spec}}
\end{figure}

\noindent{\bf \SNR(76.9)+(1.0)} (Fig.~\ref{fig:G76spec}).
This is an extended, polarised radio source in the Galactic plane, with a
non-thermal spectrum. It was first interpreted as a probable SNR by
\citet{1993A+A...276..522L}, similar to \SNR(65.7)+(1.2) ($=$DA~495, e.g.\
\citealt{1983AJ.....88.1810L}). It is $9\arcmin \times 12\arcmin$ in extent, with
diffuse emission surrounding two lobes. \citet{1993A+A...276..522L}
propose that this is a `filled-centre' remnant, with a spectral break at
relatively low frequencies, explaining their measurement of the non-thermal spectrum with
$\alpha\approx 0.62$ at higher frequencies. Subsequently
\citet{1997A+AS..123..199L} suggested that the spectral break is at $\approx
1$\,GHz. The AMI SA results are consistent with the non-thermal spectrum seen from
other observations above $\approx 1$\,GHz.
\begin{table}
\centering
\caption{Integrated flux densities from the literature
for \SNR(76.9)+(1.0).\label{tab:G76flux}}
\begin{tabular}{ccc}\hline
$\nu$/GHz & $S_{\rm{i}}$/Jy & Reference \\ \hline
 0.232 & $ 1.4\pm 0.4 $ &\citet{1997A+AS..123..199L} \\
 0.327 & $ 1.4\pm 0.1$ & \citet{1997A+AS..123..199L} \\
 0.327 & $ 2.03\pm 0.30$ & \citet{1992AJ....103..931T} \\
 0.408 & $ 2.3\pm 0.2 $ & \citet{2006A+A...457.1081K}  \\
 0.408 & $ 2.89\pm 0.20$ & \citet{1993A+A...276..522L} \\
 1.408 & $1.70\pm 0.20$ & \citet{1993A+A...276..522L} \\
 1.42 & $ 1.35\pm 0.07$ & \citet{2006A+A...457.1081K} \\
 1.49 & $1.80\pm 0.20$ & \citet{1993A+A...276..522L} \\
 2.695 & $0.90\pm 0.09$ & \citet{1993A+A...276..522L} \\
 4.80 & $0.63\pm 0.03$ & \citet{1993A+A...276..522L} \\
 4.85 & $0.44\pm 0.01$ &  \citet{1992AJ....103..931T} \\
 4.85 & $ 0.466\pm 0.028$ & \citet{1991ApJS...75.1011G} \\\hline
\end{tabular}
\end{table}

\begin{figure}
\centerline{\includegraphics[width=7.5cm,angle=-90]{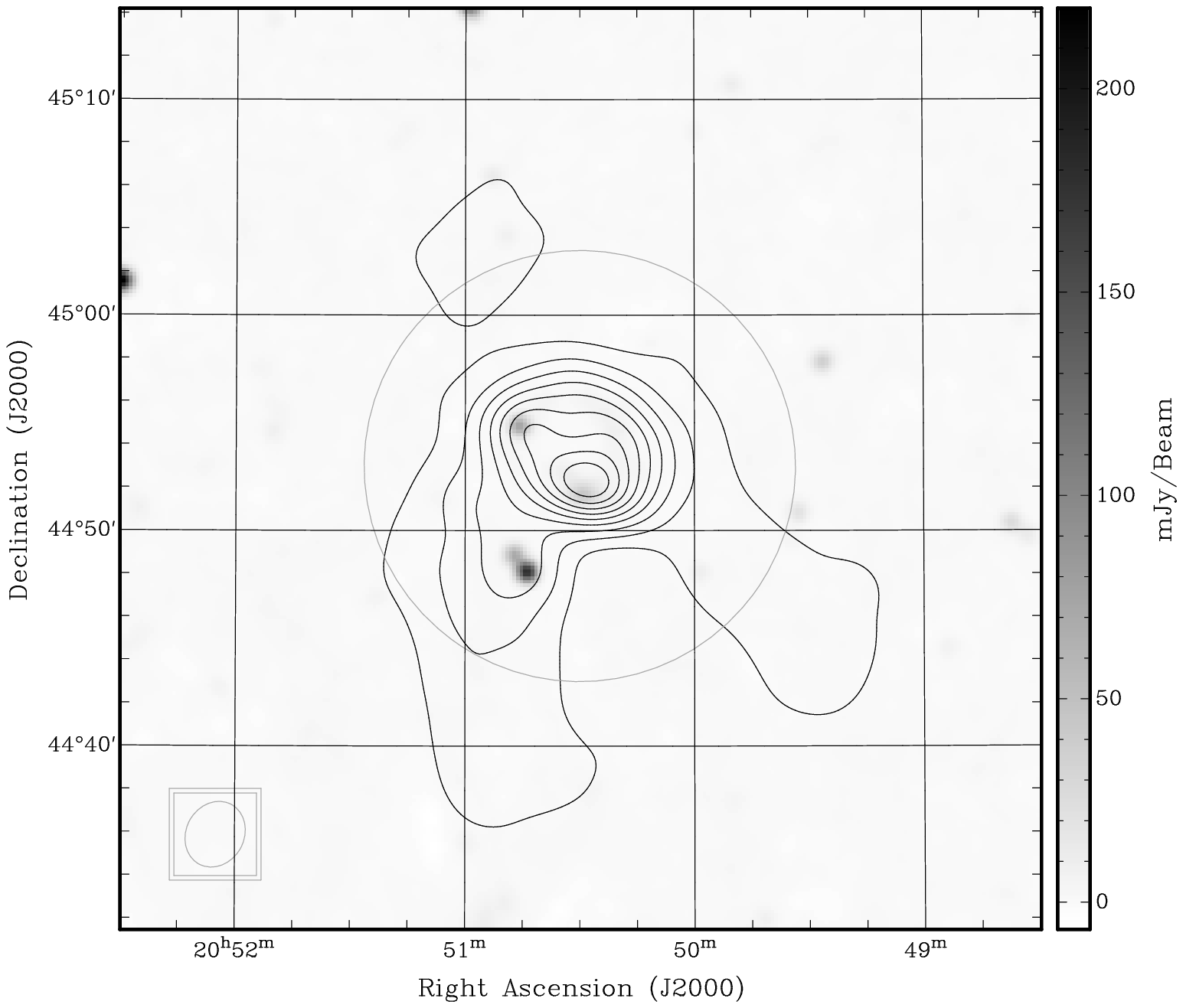}}
\vspace{0.5cm}
\centerline{\includegraphics[width=5cm,angle=-90]{G84_newspec.ps}}
\caption{Above: Map of \SNR(84.9)+(0.5). AMI SA 16-GHz contours are overlaid on
a NVSS 1.4-GHz greyscale image. Contours are as in Fig.~\ref{fig:G130spec}.
The peak level is 158\,mJy~beam$^{-1}$.
Below: Radio spectrum of \SNR(84.9)+(0.5).
Integrated flux densities taken from the literature (Table~\ref{tab:G84flux})
are shown as crosses, and those from AMI SA (Table~\ref{tab:newsnrflux})
are shown as filled circles.
The best-fitting power-law, with $\alpha=0.08$ ,is shown as a dashed line. \label{fig:G84spec}}
\end{figure}

\noindent{\bf \SNR(84.9)+(0.5)} (Fig.~\ref{fig:G84spec}).
This was identified as a SNR by \citet{1996ApJS..107..239T}, using data at
327\,MHz from the WSRT and at 4850\,MHz from \citet{1989AJ.....97.1064C}.
However, recent observations at 408 and 1420\,MHz from the CGPS
have indicated a much flatter spectrum than
previously measured. This flat spectrum is also evident across the AMI SA band (Table~\ref{tab:newsnralpha}).

Re-fitting the 327-MHz WSRT data of \citet{1996ApJS..107..239T} we find an
integrated flux density of 0.77$\pm$0.08\,Jy towards G84.9+0.5,
significantly lower than the previously measured value of $S_{\rm{i}} =
1.22\pm0.13$\,Jy \citep{1992AJ....103..931T}. This implies that
\SNR(84.9)+(0.5) is indeed an {\sc Hii} region rather than a SNR; a result which
agrees with the conclusion of \citet{2007ApJ...667..248F} based on their radio
recombination line data and also with the lack of polarized emission seen
towards this object \citep{2006A+A...457.1081K}.

\begin{table}
\centering
\caption{Integrated flux densities from the literature for
\SNR(84.9)+(0.5).\label{tab:G84flux}}
\begin{tabular}{ccc}\hline
 $\nu$/GHz & $S_{\rm{i}}$/Jy & Reference \\\hline
 0.151  & $0.917\pm 0.110$ & \citet{1998MNRAS.294..607V} \\
 0.232  & $0.63\pm 0.15$  &  \citet{1997A+AS..121...59Z} \\
 0.327  & $0.769\pm 0.077$ & This work, from the data of \\
        &                  &\citet{1992AJ....103..931T} \\
 0.408  & $0.700\pm 0.08$ & \citet{2007A+A...468..993K} \\
 1.42  &$0.77\pm  0.22$ & \citet{2007A+A...468..993K} \\
 4.85  &$0.517\pm 0.052$ & \citet{1996ApJS..103..427G} \\\hline
\end{tabular}
\end{table}

\begin{table*}
\caption{AMI SA I+Q flux densities for less well-studied SNR. Above: corrected for flux loss, where possible; below: original flux densities for those sources which undergo a flux loss correction.\label{tab:newsnrflux}}
\begin{tabular}{lcccccc}\hline
 & \multicolumn{6}{c}{Freq. (GHz)}\\ \cline{2-7}
 & 14.2 & 15.0 & 15.7 & 16.4 & 17.1 & 17.9\\
 Name  & (Jy) & (Jy) & (Jy) & (Jy) & (Jy) & (Jy)\\\hline
 \SNR(54.1)+(0.3) $^{a}$ & $0.340\pm0.034$ & $0.324\pm0.032$ & $0.281\pm0.028$ & $0.273\pm0.027$ & $0.261\pm0.026$ & $0.251\pm0.025$ \\
 \SNR(57.2)+(0.8) $^{a}$ & $0.284\pm0.033$ & $0.271\pm0.031$ & $0.252\pm0.028$ & $0.223\pm0.026$ & $0.228\pm0.026$ & $0.228\pm0.025$ \\
 \SNR(63.7)+(1.1) $^{b}$ & $0.818\pm0.082$ & $0.810\pm0.081$ & $0.790\pm0.079$ & $0.773\pm0.077$ & $0.764\pm0.076$ & $0.761\pm0.076$ \\
 \SNR(67.7)+(1.8) $^{c}$ & $0.191\pm0.019$ & $0.170\pm0.017$ & $0.169\pm0.017$ & $0.148\pm0.015$ & $0.158\pm0.015$ & $0.163\pm0.016$ \\
 \SNR(74.9)+(1.2) $^{c}$ & $2.26\pm0.23$ & $2.31\pm0.23$ & $2.26\pm0.23$ & $2.11\pm0.21$ & $2.04\pm0.20$ & $1.96\pm0.20$ \\
 \SNR(76.9)+(1.0) $^{c}$ & $0.191\pm0.019$ & $0.170\pm0.016$ & $0.170\pm0.016$ & $0.168\pm0.016$ & $0.171\pm0.016$ & $0.154\pm0.011$ \\
 \SNR(84.9)+(0.5) $^{d}$ & -- & $0.526\pm0.062$ & $0.562\pm0.058$ & $0.532\pm0.058$ & $0.540\pm0.055$ & --  \\ \hline
 \SNR(63.7)+(1.1) & $0.605\pm0.030$ & $0.591\pm0.030$ & $0.569\pm0.029$ & $0.549\pm0.028$ & $0.535\pm0.027$ & $0.525\pm0.026$ \\
 \SNR(67.7)+(1.8) & -- & -- & -- & -- & $0.148\pm0.009$ & --\\
 \SNR(74.9)+(1.2) & $1.44\pm0.09$ & $1.38\pm0.07$ & $1.36\pm0.07$ & $1.22\pm0.06$ & $1.14\pm0.06$ & $1.01\pm0.10$ \\
 \SNR(76.9)+(1.0) & $0.191\pm0.010$ & $0.162\pm0.008$ & $0.160\pm0.008$ & $0.155\pm0.008$ & $0.156\pm0.08$ & $0.142\pm0.011$ \\ \hline
\end{tabular}
\begin{minipage}{16cm}{
 $^{a}$Uncorrected for flux loss; complex region with no CGPS coverage.

 $^{b}$Corrected for flux loss using a Gaussian model.

 $^{c}$Corrected for flux loss using CGPS data.

 $^{d}$Flux loss correction smaller than other errors, so not included.
}
\end{minipage}
\end{table*}

%
%
\begin{table}
\caption{AMI SA I+Q interpolated 16\,GHz flux densities and estimated spectral indices, calculated from the AMI SA data in
Tables~\ref{tab:newsnrflux} and \ref{tab:oldsnrflux} for sources for which a flux correction has been made.\label{tab:newsnralpha}}
\begin{tabular}{ccc}\hline
 Name  &  $S_{\rm{16GHz}}$(Jy) & $\alpha$\\\hline
 \SNR(63.7)+(1.1) & $0.786^{+0.867}_{-0.412}$ & 0.34$\pm$0.24  \\
 \SNR(67.7)+(1.8) & $0.165^{+0.162}_{-0.082}$ & 0.76$\pm$0.25  \\
 \SNR(74.9)+(1.2)$^*$ & $2.189^{+1.928}_{-1.025}$ & 1.00$\pm$0.23  \\
 \SNR(76.9)+(1.0) & $0.174^{+0.123}_{-0.072}$ & 0.99$\pm$0.19  \\
 \SNR(84.9)+(0.5) & $0.533^{+0.383}_{-0.223}$ & $-0.03\pm$0.23  \\ \hline
 \SNR(120.1)+(1.4) & $10.76^{+8.96}_{-4.89}$ & 0.99$\pm$0.19  \\
 \SNR(130.7)+(3.1) & $23.97^{+14.98}_{-9.22}$ & 0.37$\pm$0.18  \\ \hline
\end{tabular}
\begin{minipage}{10cm}{
 $^*$ Fit to data above 11\,GHz.
}
\end{minipage}
\end{table}

\section{Conclusions}
New observations in six channels spanning 14.2--17.9\,GHz have improved our knowledge of the
spectral behaviour of eight supernova remnants in the northern sky.
Observations made of well-studied remnants confirm that AMI SA is measuring flux densities accurately
and our flux loss measurements and fitting methods are correct.
For those sources for which a flux loss correction was made,
Table~\ref{tab:newsnralpha} gives an interpolated flux density and spectral
index from the AMI data alone. The errors on the indices are relatively large, because of the
limited frequency coverage of the telescope, but all values are consistent within one sigma
with those calculated from the literature, except in the case of \SNR(74.9)+(1.2), as
our measurements lie above the spectral break frequency of 11\,GHz.
We detect no anomalous excess emission from the observed SNRs, implying there is no spinning dust present.
\SNR(54.1)+(0.3) shows a possible break in the
AMI SA band but, due to changes in large-scale structure over AMI SA's frequency coverage, we will need data at higher
frequencies to provide a definite answer.

\section*{Acknowledgments}

We thank the whole AMI team for their invaluable
assistance in the construction, commissioning and operation of AMI.
AMI is supported by the STFC. NHW and MLD acknowledge the support of
PPARC/STFC studentships.

\bsp \label{lastpage}


\begin{thebibliography}{1}
\setlength{\labelwidth}{0pt}
\bibitem[\protect\citeauthoryear{Aller \& Reynolds}{1985}]{1985ApJ...293L..73A} Aller H.~D. \& Reynolds S.~P.\ 1985, ApJ, 293, L73

\bibitem[\protect\citeauthoryear{Altenhoff et al.}{1979}]{1979A+AS...35...23A} Altenhoff W.~J., Downes D., Pauls T., Schraml J., 1979, A\&AS, 35, 23

\bibitem[\protect\citeauthoryear{AMI Consortium: Scaife et al.}{2007}]{2007MNRAS.377L..69S} AMI Consortium: Scaife A., et al., 2007, MNRAS, 377, L69

\bibitem[\protect\citeauthoryear{AMI Consortium: Zwart et al.}{2008}]{0807.2469} AMI Consortium: Zwart J.~T.~L. et al., 2008, MNRAS in press, pre-print (astro-ph/0807.2469)

\bibitem[\protect\citeauthoryear{Baars et al.}{1977}]{1977A+A....61...99B} Baars J.~W.~M., Genzel R., Pauliny-Toth I.~I.~K., Witzel A., 1977, A\&A, 61, 99

\bibitem[\protect\citeauthoryear{Becker, White \& Edwards}{1991}]{1991ApJS...75....1B} Becker R.~H., White R.~L., Edwards A.~L., 1991, ApJS, 75, 1

\bibitem[\protect\citeauthoryear{Bock \& Gaensler}{2005}]{2005ApJ...626..343B} Bock D.~C.-J., Gaensler B.~M., 2005, ApJ, 626, 343

\bibitem[\protect\citeauthoryear{Boland et al.}{1966}]{1966ApJ...144..437B} Boland J.~W., Hollinger J.~P., Mayer C.~H., McCullough T.~P.\ 1966, ApJ, 144, 437

\bibitem[\protect\citeauthoryear{Browne et al.}{1998}]{1998MNRAS.293..257B} Browne I.~W.~A., Wilkinson P.~N., Patnaik A.~R., Wrobel J.~M., 1998, MNRAS, 293, 257

\bibitem[\protect\citeauthoryear{Caswell \& Haynes}{1987}]{1987A+A...171..261C} Caswell J.~L., Haynes R.~F., 1987, A\&A, 171, 261

\bibitem[\protect\citeauthoryear{Chini, Kruegel \& Wargau}{1987}]{1987A+A...181..378C} Chini R., Kruegel E., Wargau W., 1987, A\&A, 181, 378

\bibitem[\protect\citeauthoryear{Condon, Broderick \& Seielstad}{1989}]{1989AJ.....97.1064C} Condon J.~J., Broderick J.~J., Seielstad G.~A., 1989, AJ, 97, 1064

\bibitem[\protect\citeauthoryear{Condon et al.}{1998}]{1998AJ....115.1693C} Condon J.~J., Cotton W.~D., Greisen E.~W., Yin Q.~F., Perley R.~A., Taylor G.~B., Broderick J.~J., 1998, AJ, 115, 1693

\bibitem[\protect\citeauthoryear{Duin et al.}{1975}]{1975A+A....38..461D} Duin R.~M., Israel F.~P., Dickel J.~R., Seaquist E.~R., 1975, A\&A, 38, 461

\bibitem[\protect\citeauthoryear{Fesen}{2008}]{2008ApJS..174..379F} Fesen R. et al., 2008, ApJS, 174, 379F

\bibitem[\protect\citeauthoryear{Foster et al.}{2007}]{2007ApJ...667..248F} Foster T.~J., Kothes R., Kerton C.~R., Arvidsson K., 2007, ApJ, 667, 248

\bibitem[\protect\citeauthoryear{F\"urst et al.}{1990}]{1990A+AS...85..805F} F\"urst E., Reich W., Reich P., Reif K., 1990, A\&AS, 85, 805

\bibitem[\protect\citeauthoryear{Geldzahler, Pauls \& Salter}{1980}]{1980A+A....84..237G} Geldzahler B.~J., Pauls T., Salter C.~J., 1980, A\&A, 84, 237

\bibitem[\protect\citeauthoryear{Gotthelf et al.}{2007}]{2007ApJ...654..267G} Gotthelf et al, 2007 ApJ, 654, 267G

\bibitem[\protect\citeauthoryear{Green}{1986}]{1986MNRAS.218..533G} Green D.~A., 1986, MNRAS, 218, 533

\bibitem[\protect\citeauthoryear{Green}{2004}]{2004BASI...32..335G} Green D.~A., 2004, BASI, 32, 335

\bibitem[\protect\citeauthoryear{Green}{2007}]{2007BASI...35...77G} Green D.~A., 2007, BASI, 35, 77

\bibitem[\protect\citeauthoryear{Green, Baker \& Landecker}{1975}]{1975A+A....44..187G} Green A.~J., Baker J.~R., Landecker T.~L., 1975, A\&A, 44, 187

\bibitem[\protect\citeauthoryear{Green \& Scheuer}{1992}]{1992MNRAS.258..833G} Green D.~A., Scheuer P.~A.~G., 1992, MNRAS, 258, 833

\bibitem[\protect\citeauthoryear{Green, Tuffs \& Popescu}{2004}]{2004MNRAS.355.1315G} Green D.~A., Tuffs R.~J., Popescu C.~C., 2004, MNRAS, 355, 1315

\bibitem[\protect\citeauthoryear{Gregory \& Condon}{1991}]{1991ApJS...75.1011G} Gregory P.~C., Condon J.~J., 1991, ApJS, 75, 1011

\bibitem[\protect\citeauthoryear{Gregory et al.}{1996}]{1996ApJS..103..427G} Gregory P.~C., Scott W.~K., Douglas K., Condon J.~J., 1996, ApJS, 103, 427

\bibitem[\protect\citeauthoryear{Griffith et al.}{1990}]{1990ApJS...74..129G} Griffith M., Langston G., Heflin M., Conner S., Lehar J., Burke B., 1990, ApJS, 74, 129

\bibitem[\protect\citeauthoryear{Handa et al.}{1987}]{1987PASJ...39..709H} Handa T., Sofue Y., Nakai N., Hirabayashi H., Inoue M., 1987, PASJ, 39, 709

\bibitem[\protect\citeauthoryear{Katz-Stone et al.}{2000}]{2000ApJ...529..453K} Katz-Stone D.~M., Kassim N.~E., Lazio T.~J.~W., O'Donnell R., 2000, ApJ, 529, 453

\bibitem[\protect\citeauthoryear{Klein et al.}{1979}]{1979A+A....76..120K} Klein U., Emerson D.~T., Haslam C.~G.~T., Salter C.~J., 1979, A\&A, 76, 120

\bibitem[\protect\citeauthoryear{Kneissl et al.}{2001}]{2001MNRAS.328..783K} Kneissl R., Jones M.~E., Saunders R., Eke V.~R., Lasenby A.~N., Grainge K., Cotter G., 2001, MNRAS, 328, 783

\bibitem[\protect\citeauthoryear{Kneissl \& Jones}{2002}]{2002ASPC..268..121K} Kneissl R., Jones M.~E., 2002, ASPC, 268, 121

\bibitem[\protect\citeauthoryear{Kothes et al.}{2006}]{2006A+A...457.1081K} Kothes R., Fedotov K., Foster T.~J., \& Uyan{\i}ker B.\ 2006, AAp, 457, 1081

\bibitem[\protect\citeauthoryear{Kothes \& Dougherty}{2007}]{2007A+A...468..993K} Kothes R., Dougherty S.~M., 2007, A\&A, 468, 993

\bibitem[\protect\citeauthoryear{Kovalenko, Pynzar \& Udal'tsov}{1994}]{1994ARep...38...95K} Kovalenko A.~V., Pynzar A.~V., Udal'tsov V.~A., 1994, ARep, 38, 95

\bibitem[\protect\citeauthoryear{Landecker \& Caswell}{1983}]{1983AJ.....88.1810L} Landecker T.~L., Caswell J.~L., 1983, AJ, 88, 1810

\bibitem[\protect\citeauthoryear{Landecker, Higgs \& Wendker}{1993}]{1993A+A...276..522L} Landecker T.~L., Higgs L.~A., Wendker H.~J., 1993, A\&A, 276, 522

\bibitem[\protect\citeauthoryear{Landecker et al.}{1997}]{1997A+AS..123..199L} Landecker T.~L., Zheng Y., Zhang X., Higgs L.~A., 1997, A\&AS, 123, 199

\bibitem[\protect\citeauthoryear{Langston et al.}{2000}]{2000AJ....119.2801L} Langston G., Minter A., D'Addario L., Eberhardt K., Koski K., Zuber J., 2000, AJ, 119, 2801

\bibitem[\protect\citeauthoryear{Lockman}{1989}]{1989ApJS...71..469L} Lockman F.~J.\ 1989, ApJS, 71, 469

\bibitem[\protect\citeauthoryear{Mason et al.}{1999}]{1999AJ....118.2908M} Mason B.~S., Leitch E.~M., Myers S.~T., Cartwright J.~K., Readhead A.~C.~S., 1999, AJ, 118, 2908

\bibitem[\protect\citeauthoryear{Metropolis et al.}{1953}]{1953JoCP...21.1087} Metropolis N., Rosenbluth A.~W., Rosenbluth M.~N., Teller E., 1953, Journal of Chemical Physics, 21, 1087-1092

\bibitem[\protect\citeauthoryear{Morsi \& Reich}{1987}]{1987A+AS...71..189M} Morsi H.~W., Reich W., 1987, A\&AS, 71, 189

\bibitem[\protect\citeauthoryear{O'Sullivan \& Green}{1999}]{1999MNRAS.303..575O} O'Sullivan C., Green D.~A., 1999, MNRAS, 303, 575

\bibitem[\protect\citeauthoryear{Patnaik et al.}{1992}]{1992MNRAS.254..655P} Patnaik A.~R., Browne I.~W.~A., Wilkinson P.~N., Wrobel J.~M., 1992, MNRAS, 254, 655

\bibitem[\protect\citeauthoryear{Pineault \& Chastenay}{1990}]{1990MNRAS.246..169P} Pineault S., Chastenay P., 1990, MNRAS, 246, 169

\bibitem[\protect\citeauthoryear{Rees}{1990}]{1990MNRAS.243..637R} Rees N., 1990, MNRAS, 243, 637

\bibitem[\protect\citeauthoryear{Reich et al.}{1984}]{1984A+AS...58..197R} Reich W., F\"urst E., Haslam C.~G.~T., Steffen P., Reif K., 1984, A\&AS, 58, 197

\bibitem[\protect\citeauthoryear{Reich et al.}{1985}]{1985A+A...151L..10R} Reich W., F\"urst E., Altenhoff W.~J., Reich P., Junkes N., 1985, A\&A, 151, L10

\bibitem[\protect\citeauthoryear{Reynolds \& Ellison}{1992}]{1992ApJ...399L..75R} Reynolds S.~P., Ellison D.~C., 1992, ApJ, 399, L75

\bibitem[\protect\citeauthoryear{Salter et al.}{1989}]{1989ApJ...338..171S} Salter C.~J., Reynolds S.~P., Hogg D.~E., Payne J.~M., Rhodes P.~J., 1989, ApJ, 338, 171

\bibitem[\protect\citeauthoryear{Sieber \& Seiradakis}{1984}]{1984A+A...130..257S} Sieber W., Seiradakis J.~H., 1984, A\&A, 130, 257

\bibitem[\protect\citeauthoryear{Stephenson \& Green}{2002}]{2002hsr..book.....S} Stephenson F.~R., Green D.~A., 2002, ``Historical Supernovae and their Remnants'', (OUP)

\bibitem[\protect\citeauthoryear{Taylor et al.}{1996}]{1996ApJS..107..239T} Taylor A.~R., Goss W.~M., Coleman P.~H., van Leeuwen J., Wallace B.~J., 1996, ApJS, 107, 239

\bibitem[\protect\citeauthoryear{Taylor, Wallace \& Goss}{1992}]{1992AJ....103..931T} Taylor A.~R., Wallace B.~J., Goss W.~M., 1992, AJ, 103, 931

\bibitem[\protect\citeauthoryear{Tian \& Leahy}{2005}]{2005A+A...436..187T} Tian W.~W., Leahy D.~A., 2005, A\&A, 436, 187

\bibitem[\protect\citeauthoryear{Tian \& Leahy}{2006}]{2006A+A...451..991T} Tian W.~W., Leahy D.~A., 2006, A\&A, 451, 991

\bibitem[\protect\citeauthoryear{Trushkin, Vitkoskij \& Nizhelskij}{1987}]{1987AISAO..25...84T} Trushkin S.~A., Vitkoskij V.~V., Nizhelskij N.~A., 1987, AISAO, 25, 84

\bibitem[\protect\citeauthoryear{Trushkin}{1996}]{1996BSAO...41...64T} Trushkin S.~A., 1996, BSAO, 41, 64

\bibitem[\protect\citeauthoryear{Uro{\v s}evi{\'c}, Pannuti \& Leahy}{2007}]{2007ApJ...655L..41U} Uro{\v s}evi{\'c} D., Pannuti T.~G., Leahy D., 2007, ApJ, 655, L41

\bibitem[\protect\citeauthoryear{Velusamy \& Becker}{1988}]{1988AJ.....95.1162V} Velusamy T., Becker R.~H., 1988, AJ, 95, 1162

\bibitem[\protect\citeauthoryear{Vessey \& Green}{1998}]{1998MNRAS.294..607V} Vessey S.~J. \& Green D.~A. 1998, MNRAS, 294, 607

\bibitem[\protect\citeauthoryear{Vinyaikin}{2007}]{2007ARep...51..570V} Vinyaikin E.~N., 2007, ARep, 51, 570

\bibitem[\protect\citeauthoryear{Wallace, Landecker \& Taylor}{1997}]{1997AJ....114.2068W} Wallace B.~J., Landecker T.~L., Taylor A.~R., 1997, AJ, 114, 2068

\bibitem[\protect\citeauthoryear{Weiler \& Shaver}{1978}]{1978A+A....70..389W} Weiler K.~W., Shaver P.~A., 1978, A\&A, 70, 389

\bibitem[\protect\citeauthoryear{Wendker, Higgs \& Landecker}{1991}]{1991A+A...241..551W} Wendker H.~J., Higgs L.~A., Landecker T.~L., 1991, A\&A, 241, 551

\bibitem[\protect\citeauthoryear{White \& Becker}{1992}]{1992ApJS...79..331W} White R.~L., Becker R.~H., 1992, ApJS, 79, 331

\bibitem[\protect\citeauthoryear{Wilkinson et al.}{1998}]{1998MNRAS.300..790W} Wilkinson P.~N., Browne I.~W.~A., Patnaik A.~R., Wrobel J.~M., Sorathia B., 1998, MNRAS, 300, 790

\bibitem[\protect\citeauthoryear{Zhang et al.}{1997}]{1997A+AS..121...59Z} Zhang X., Zheng Y., Chen H., Wang S., Cao A., Peng B., Nan R.\ 1997, A\&AS, 121, 59

\end{thebibliography}
\end{document}